# Base of nonlinear dynamics
# Or Real Dynamics, Ideal Dynamics, Unpredictable Dynamics and " Schrodinger cat ".

**Kupervasser Oleg**

**Abstract.**

In the paper paradoxes underlying thermodynamics and a quantum mechanics are discussed. Their solution is given from the point of view of influence of the exterior observer (surrounding medium) destroying correlations of system, or boundedness of self-knowledge of system in a case when both the observer, and a surrounding medium are included in system. Concepts Real Dynamics, Ideal Dynamics and Unpredictable Dynamics are entered. Consideration an appearance of a life is given from the point of view of these three Dynamics.

## Introduction.

Statistical mechanics and quantum mechanics - two designed and well-known theories, a basis of a modern physics. Nevertheless they contain a series of paradoxes which force many scientists to doubt about interior consistency of these theories.

We list them:

1) The basis equations classical and a quantum mechanics reversible in time while laws of thermodynamics, for example, a law of increase entropy in closed systems,it is nonreversible, though this law should be output from these equations [1], [2].
2) the similar problem can be seen in an inconsistency between the law of increase entropy in the isolated systems and the law of Poincare about returns of the isolated system as much as close to an initial state [1], [2].
3) Paradox "Schrodinger cat ", i.e. a reduction of wave package and transition from pure to the mixed state in macroscopic systems or during measuring quantum system. This process is not featured by the equations of a quantum mechanics, it is completely not clear in what moments of time it there is also what its duration. Nevertheless, without it the quantum mechanics is incomplete [3].

4) The microscopic entropy of classical systems, defined through a density function in a phase space, remains constant. (it is similar to an entropy of quantum systems, defined through a density matrix [15]) The macroscopic entropy of systems, defined through macroscopic variable or, that mathimatically is equivalent, the density function "coarsened" in a phase space for classical mechanics(and the reduced density matrix for a quantum mechanics), can both to increase and to decrease with identical probability because of reversibility of a motion. (As against very popular belief, that processes with decrease of an entropy are less probable. Actually, states with a constant and peak entropy for the given system are most probable, whereas the states leading as to decrease and to increase of an entropy are equally rare in comparison with these equilibrium states) Why, in that case, in the real world in all macroscopic classical systems only increase of an entropy always is observed? Why to receive laws for this increase of an entropy it is necessary to intreduce certain additional guesses (like a Bolzmann hypothesis of the molecular chaos or the equivalent to it on sense "coarsened" density functions in a phase space), besides the basic equations of classical mechanics though they and give us complete exposition of reality? [1], [2]

5) For exposition of a macroscopic state are use some macroscopic variable. Theoretically their choice is restricted to nothing, it can be any function of microscopic parameters, nevertheless, as a rule, choice "convenient" variable It is predetermined enough. Than?

6) There is a set of alive systems, including possessing consciousness and freedom of will. What physical properties in essence distinguish alive (understanding) systems from lifeless? As far as it is full they can be described by the known equations of physics and thermodynamics?

it is not by chance that Schrodinger used for a statement of the paradox as the macroscopic detector of an alive cat! A certain connection between the basis paradoxes of physics described above and secrets of a life intuitively is clear to many physicists. It is necessary to try to formulate this connection clearly.

The solution of these paradoxes does not demand the invention of new laws of physics though their experimental checkout can lead to to them. We shall analyze these problems strictly within the framework of known laws of physics.

There is an extensive literature devoted to these paradoxes [1], [2], [3]. Much from made is correctly. A problem that these viewings do not give the key property of dynamic system giving in all these paradoxes and give only fragments of their complete solution.

The situation here reminds paradox of Gibbs [4] where jump of an entropy at an intermixture of gases explained absence of the transition shapes between identical and different gases. Nevertheless such transition shapes are possible, quantity of the jump can be controled and consideration (not trivial) gives not only the deep understanding and the solution of this paradox, but also more the deep understanding of base of physics and does not demand introduction of new laws.

The similar situation developed and with a spin of an electron: textbooks are overflown with statements, that it cannot be interpreted as the natural moment of physical (i.e. in space) gyrations of a particle. Funny, nevertheless, that within the framework of the equation of Dirac it is perfectly done also a spin can be interpreted as the natural moment of only physical gyration of a wave function [5].

As well the above-stated paradoxes can be deeply understood and give by accurate viewing a rich material for understanding of base of existing physics with restrictions proper in it for an explanation of base of world around. The physics is not so omnipotent even only theoretically (as almost everything, even professional physics, are sure) and its these restrictions can be deduced quite from itself, not devising new theories. These restrictions also conduct to the mentioned above paradoxes. That is amusing, at all their variety and an exterior disconnectedness all of them have a uniform root and the reason.!

The tool for our viewing will be Nonlinear Dynamics, a science which doesn't sets as itself the purpose the invention of new fundamental laws of physics, and is used for more deep understanding already existing laws and for a determination of blanket properties and laws concerning to completely different physical systems as it would seem (in sense of a subject domain of their application), but, nevertheless, having very similar dynamics.

## Part 1. Quantum paradox « Schrodinger cat ».

Here we shall not feature explicitly a requirement of this paradox, they are well described in many sourses[14], [3], [6], [7]. We shall stop here on its essence. The cat is a macroscopic system which, being is described macroscopicly (quantumly), can is simultaneously in two states - a cat alive and a cat dead, being them « the weighed sum » (so-called a wave packet of psi functions)! In a reality, both the possible outside observer, and a cat consciousness fix always only one of these two states with probability, defined quadrates of their "weights" in above described to " the weighed sum ». In language of mathematics it corresponds to transition from pure to the mixed state and a diagonalization of a density matrix.

This process is named of a reduction of a wave packet and is not featured by the equations of a quantum mechanics. Moreover, it corresponds to increase of a microscopic entropy (defined with the help of density matrix ) coordinating thus with the previous two paradoxes and conflicting with reversibility of the quantum equations and "returns" of quantum system. As these paradoxes are proper also in classical mechanics It will allow us to find further as well classical analog for paradox « Schrodinger cat » which perfectly exists, despite of universal complete conviction in its " only quantum nature ", connected with lack of the severe understanding of essence and the reasons of these paradoxes. Necessity of introduction of this mysterious reduction of a wave packet for macroscopic systems (being, as a rule, last objects in measuring systems, i.e. their detectors) is a basis and essence of paradox of « Schrodinger cat ».

By the way, in the literature there are attempts to describe by help of art certain states of consciousness adequate to quantum " weighed sum " [8]. But if those also exist actually, they are more likely exotic, than a rule.

For the solution of paradox, apparently, it would be easiest to assume, that evolution primely breaks up on two possible « branches of evolution » with particular probability everyone, is similar to how it happens in statistical mechanics where the state of system is featured not by a point in a phase space, and is certain "cloud" of such points. The matter is that a reduction of a wave packet, as against « partitionings into branches » evolutions in statistical mechanics, leads to other evolution for branches, than it would be,if we use « the weighed sum ». I.e. these two possible branches of evolution render cross influence against each other through mixture of effects of final measuring (which is reduction), because « the weighed sum» solution is arithmetical sum of these branches, instead of a statistical intermixture of these branches.

It can be confirmed in the following examples. The first example is connected to reversibility of quantum equations of motion. As it has already been told, the reduction of a wave packet leads to to increase of an entropy while reversible quantum evolution leaves its constant.

It havn't sense to think that these two, so different, evolution types can give identical results.

The second example is connected to the Poincare theorem about "returns" of dynamic system [1], [2]. It would be desirable to accent here on some characteristics of this theorem for quantum systems. In case of classical mechanics the majority of dynamic systems are chaotic: return times form casual sequence and strongly vary on quantity at the slightest modifications of starting conditions. Nevertheless there is though also small, but physically significant class of systems with periodic or nearly so periodic return times,stsble against small errors of initial conditions. These systems are integrable in variables action - angle.

In case of quantum systems a situation directly inverse - the isoleted system with finite volume and with a finite number of particles - always possesses periodic or nearly periodic return times, except for systems of the infinite volume or the infinite number of particles where this return time is equal to infinity [9].

Accordingly, being returned to our equations, the reduction of the cat leads to to increase of a microscopic entropy, doing impossible return because it corresponds to previous, smaller entropy, but during usual reversible quantum evolution the entropy is constant. Thus, the reduced and not reduced systems have various dynamics.

The third amusing example - "kettle which never will begin to boil" [3]. Let we have a particle which should pass from upper to the down level of energy in correspondence with laws of a quantum mechanics (for example to break up). If to carry out observations over this system too frequently it never will make! Changes of evolution of system by a reduction happening at the moment of observation, leads to to this effect. How frequently there are these acts of a reduction in a reality and what their natural time duration? it is not clear.

By the way, this paradox explains, why decay of particles or states in a quantum mechanics is not precisely exponential, and only to the close to it. These diversions, apparently, allow to define "age" of system as only exponential type of decay makes it impossible. Nevertheless it not process of observation and contortions intreduced by it because of a reduction do an differennce between exponential and quantum type of decay unobservable.

These reasonings prove, that the reduction is not primely mechanical partitioning of a motion into two possible independent branches, it changes all actual dynamics of system.

Let's consider possible known approaches to the solution of paradox of a cat. All these approaches represent certain exact steps in the necessary direction, but do not lead to to the complete solution of paradox.

Last time has appeared a huge amount of various interpretations of the quantum mechanics, the most known and popular of which "multiuniversal"[3] which guesses, that at the moment of a reduction of a wave packet is not picked only one branch, and they exist simultaneously in certain parallel words, in each of these worlds the observer sees only one of branches, nevertheless, as they exist simultaneously when measuring will be carried out, all possible "worlds" influence effects of this measuring, ensuring the cross influence of branches arising from laws of a quantum mechanics which at a reference reduction is destroyed by a select only one of branches. The problem of this interpretation (however as well as all others) that it does not resolve difficulties, and only transfers them to other, less noticeable for us, a plane. So, the observer in any of possible "worlds" has the complete information only about him and anything in the explicit shape does not know about other worlds that does actually his information inexact and not full for a prediction of effects of measuring which are determined by final evolution in all "worlds". Thus, violation of normal quantum evolution in case of a reduction is exchanged by the principal unpredictability of results of measuring because of the restricted information in each of "worlds". Though the hint on the exact solution of paradox of cat here is.

The following approach to the solution of paradox of a cat is the account of influence of the exterior observer. In fact the reduction of a wave packet also happens at the moment of observation or measuring. Really, the reduction and violation of a normal course of quantum evolution can be explained in this case by the fact that systems is unisolated and consequently by action of exterior forces.

Here to us follows will stop on one of the main principal moment distinguishing classical and quantum systems. In classical systems influence of the measuring device on explored process can be (certainly only theoretically, instead of practically) is made as much as small. In a quantum mechanics, measuring cannot be carried out without a reduction of a wave function and though also small, but finite final action on measured system. Than more precisely we want to measure a state of quantum system, then so stronger perturbation we we must import in it. It does a quantum mechanics in essence not closed. In it, but not in other type of the basis equations or probability character of laws as it is thinking usually, the key difference of a quantum and classical mechanics will consist. It is necessary to note especially, that it at all does not do impossible precise reversible quantum description of systems or their experimental checkout by the exterior observer. Primely it is possible in a restricted cases, only under specific requirements which we shall touch below. Here only we shall mark, that checkout by the exterior observer is reversible quantum equations of motion correct is possible, but not a precise measure of parameters and histories of evolution of system, becausse them are featured not only by the equations, but also the initial state of system defenitely broken by observation.

The impossibility to exclude influence of measuring is illustrated by experience with the quantum achieving on the screen with two slots, and an interference pattern behind it: at measuring through what from two slots has passed quantum the interference patern disappears, being exchanged by the simple sum of intensities of signals from two slots. For giving to process of interaction with the observer of objectivity and independence of the subjective observer his action can be exchanged by uncontrollable interaction with a macroscopic surrounding medium, that also make system to be not isolated. This process gets a title decoherentization [16] and has the deserved popularity. Its main advantage will be, that it exchanges a set of quantum reversible equations of motion for the isolated system and a mysterious irreversible process of a reduction by the closed set of equations including exterior noise. These equations have concrete and practical meaning. Here we are waited also with new difficulties. Less important difficulty will be, that parameters of system and an equation of motion become dependent from some uncontrollable exterior noise. What this noise should be chosen to provide correct and exact description of system?

Much more principal difficulty will be, that both the observer, and an environment can be included in system and process of a reduction which is really observed and does not disappear anywhere in this case, will already happen in the isolated system. In this case, the explanation of a reduction found above as result of exterior action, disappears so we are returned to our paradox in its former unresolved shape. The elementary example for understanding of this appearance is the case when the cat is engaged in introspection, stating, that he is still alive, instead of dead. In this case a final stage of measuring will be consciousness of a cat. What to do in this case? To insert the infinite sequential line-up of observers, increasing our system ad infinitum? To assign to our consciousness actual material force and ability to a reduction of system or even how it is sometimes offered, to create a certain new physics of consciousness [7]? The real solution is much easier and more beautiful.

In a case when the cat is under exterior observation and the observer or medium are not included in system as we saw, the paradox does not appear. A problem here will be precisely to find a concrete possible type of exterior action that is precisely impossible because acting objects are not included in consideration. What to do when all exterior factors are included
to system, including when the observer will be a cat himself? Evolution and dynamics of system will be precisely various for a case of reversible dynamics without a reduction and nonreversible with a reduction. Most brightly it is illustrated by absence of return of system to an initial state at presence of a reduction because of increase of an entropy in this process and returns, occuring for almost-periodic motion, predicted by the reversible equations of a quantum mechanics. The solution of this paradox consists that this difference between two dynamics though really exists, but cannot be tested in practice, i.e. observationally. Really, as the isolated system cannot itself exact and univalent measure its complete state or solve itself full set its equations featuring its dynamics, it is not capable also completely observationally to test its dynamics , to make the unique choice between these two possible dynamics, because they give identical result in the field where they can be chacked.
So a system cannot test own return to an initial state because to make this checkout, it should remember, that before already was in this state some time back, but this "memory" should have a certain material carrier which could not exist in an initial state in any way in which the system as we assume has returned as in this case and "memory" is returned in a former state, thus forgetting all "history". I.e., having returned in a former state, the system should lose inevitably memory of all history and cannot register return. Only the exterior independent observer with independent exterior storage is capable on similar checkout. He (or an environment instead of him) also will bring in inevitable for quantum mechanics errors to measured system which will explain a reduction. Thus, presence of two various dynamics for the same system can be explained by impossibility observationally to find differences between them, instead of mysterious creative forces of consciousness.

Here it is important to note, that a basis of this paradox and the reason, that it was so explicitly exhibited in a quantum mechanics, that just in a quantum mechanics measuring inevitably gives small, but existing perturbation of measured system. Presence of similar perturbation, but not its probability character, also distinguishes in essence a quantum mechanics from classical, leading to to very difficult paradoxes.

Here it is necessary to note once again, that, nevertheless, under strictly particular requirements which will be circumscribed below, precise checkout of reversible quantum laws is possible, despite of this restriction.

**Part 2. Classical analog of quantum paradox of «Schrodinger cat».**

The paradox of « Schrodinger cat », as a rule, is considered, as only quantum, not having analog in classical mechanics. It is , nevertheless, mistake. The analog a reduction of a wave packet takes place and in classical mechanics! Really, the law increase of an entropy (in a quantum mechanics happening in the moment a reduction of a wave packet) happens and in classical statistical mechanics, and by analogy to a quantum mechanics, enters an inconsistency with reversible character of laws of a motion and the theorem of Poincare about returns. These classical laws of a motion leave an entropy constant (if it is defined through a density function in a phase space), or it can both to increase, and to decrease (if it is defined through the "coarsened" density function in a phase space, i.e. averaged in some neighbourhood of each point of a density function in a phase space). How to explain only increasing of an entropy in a reality? Introduction Bolzmann « hypotheses of the molecular chaos » [1], [2] which also is one of types "coarsening" frequency functions in a phase space, result in losses of correlations between velocities and positions of different molecules (that expresses in reversibility of a motion at opposite velocities and returns through known time in a small neighbourhood of an initial state) and nonreversible evolution of system. But we here it visual the complete analogy to a reduction of a wave packet! It also leads to to losses of correlations (nondiagonal elements of a density matrix) and nonreversibility of a motion! I.e. « The hypothesis of the molecular chaos » or other type "coarsening" of density functions in a phase space also is the complete analog of a reduction in a quantum mechanics. Losses of correlations at these operations it is equivalent to zeroing of nondiagonal elements of a density matrix at a reduction of a wave packet. What is possible explain to introduction of these additional to classical mechanics and contradicting to classical mechanics "coarsening"? The same reasons, as explanation of reduction in a quantum mechanics: an indistinguishability of two dynamics (with and without "coarsening") at real experiments (for the isolated systems because of impossibility of "memorization" of starting conditions at introspection because of returns, for exterior observation by interaction of the observer or an environment with apparent system). But here we meet with main and key distinction of a classical and quantum mechanics.

If interaction of the observer with observed system always is present at a quantum mechanics in classical mechanics it can be made theoretically to zero! In practice it always takes place. It can explain the inconsistency between theoretical observability of decrease of an entropy and its absence in a reality in large systems: real, finite and small interaction with the observer or primely "environment" leads to destroying of processes with decrease of an entropy. Really, processes with decrease of an entropy, as against processes with increase of an entropy, are strongly unstable in relation to random exterior noise that leads to to their destoying and sync of deflections of time arrows between the observer, observed system and their environment, even for small interaction between them. The positive direction of time arrow is defined in a direction of increase of an entropy. By introduction of these positive direction of time arrow it tried to explain earlier a law of degradation of energy. Thus there was a problem: as both directions of a deflection of time are equal probability (only equilibrium states without explicitly expressed direction are maximum probable) why in a reality all these arrow of time directions are the same? It was considered as a unsolvable problem for which solution it is necessary to search almost in depth of an origin of the Universe [28] while the solution is very simple and lays only in real, finite and small interaction of all subsystems resulting in universal sync of all deflections of time. It is necessary to note, that the theoretical possibility of zero interaction between systems in classical mechanics has led to to that all these appearances, including paradox «Schrodinger cat » were exhibited in the explicit and precise shape only within the framework of a quantum mechanics while they are characteristic also for classical mechanics for small, but finite interaction between systems which always takes place in the real world, except for some very thin and not natural situations created by people in experiments. To these situations we now also shall transfer. Before only we shall agree, that in the further text we call to system

"real",if such interaction, though small with an environment or the observer presences for system though in classical mechanics it can be done zero.

**Part 3. The experimental observation of quantum paradox «Schrodinger cat» and the experimental checkout of Ideal dynamics.**

In classical mechanics such situation forms very simply: almost ideal isolation of system from an exterior environment and introduction of almost zero interaction of system with the observer. In a quantum mechanics it is impossible: measuring of system always is connected to inherent interaction. It can seem, that difference between reduced or not reduced dynamics (we shall be not reduced dynamics (in classical mechanics dynamics before introduction "coarsening") further to term as Ideal Dynamics) in any case becomes really not checked. This deduction, nevertheless, is erroneous. Measuring in a quantum mechanics is possible two types: when the result of measuring corresponds to a state of system before or after measuring from theoretically 100 % precision. Both states cannot be simultaneously measured because of interaction with the observer. And the measuring relevant to a state of system after measuring is more likely not real measuring, but "preparation" of system for measuring because the initial state changes and remains to unknowns. The measuring relevant to a state of system after measuring we shall name "observation". Thus, quantum system with "preparation" during the initial moment, almost the complete isolation in the intermediate state from observation and an environment an "observation" in a final state is a polygon for checkout of Ideal dynamics and a quantum equivalent of classical isolated systems.
It has the following essential deficiencies:

1) The initial state varies and remains unknowns.
2) The intermediate states are not measured and remain unknown. As a matter of fact it is possible to compare only initial and final states on correspondence to Ideal dynamics.
3) Cases when it is possible to achieve so the complete isolation are rare and demand huge efforts.

Examples of such systems:

1) Mesoscopical systems at low temperatures [10]. These systems due to enogh large sizes are close to boundary of applicability of the statistical law of large numbers and almost macroscopic. Due to low temperatures (and, accordingly, momentums and velocities of molecules) sizes of a quantum wave packet are great (because of an indeterminacy principle of quantum mechanics), are close to a size of system and support, accordingly, quantum correlations. Interaction with an environment is small and its quantity is easily controllable. Thus, it is possible to check quantum coherent oscillations or tunnel effect on rather large, almost macroscopic systems. All experiments carried out till now confirm realization of Ideal dynamics, instead of a reduction in the intermediate states.

2) Systems near to phase transitions of II sort. The such systems have long size correlations, comparable with sizes of system.

3) Perhaps, any types of alive processes or their primitive possible prototypes.

**Part 4. Definition of Real Dynamics, Ideal Dynamics, Unpredictable Dynamics and the Macroscopic State. Prigogine « New Dynamics » .**

Ideal Dynamics we have agreed to term standard equations of a quantum and classical mechanics. In a reality, except for a small amount of the cases described above, because of impossibility of the complete self-observation of system or inherent Interactions with the exterior observer (medium) Ideal Dynamics observationally is unchecked and behaviour of system becomes, strictly speaking, not predicted. There is Unpredictable Dynamics instead of Ideal Dynamics.

In practice, however, the majority of systems is well featured and predicted by laws of physics. How it is possible?!

There are two major factors, giving to unpredictability:

1) Impossibility of the complete self-observation if the observer and medium are included in system of observation. It superimposes restriction on precise knowledge of starting conditions of a motion.

2) Uncontrollable interaction of the exterior observer or an environment with featured system. It superimposes restriction on precise knowledge of equations of motion because of uncontrollable exterior noise.

There is, however, a solution of these difficulties. It will consist in replacement of the complete exposition of system on reduced through introduction macroscopic variable, being certain functions microvariable. Thus here this concept is interpreted very widely, for example, the knowledge of velocities and positions of all molecules with anyone, but finite precision is also macroexposition of system.

The startling fact, but for major number of real systems almost always exists (and not one!) a set of macrovariables at which the equations of their motion become in very wide diapason of exterior noise or errors of initial conditions INDEPENDENT (or nearly so INDEPENDENT) from quantity and a concrete form of these noise or errors of initial conditions during a time interval smaller then half of recovery time for periodic or nearly periodic systems and even to the infinite for chaos systems with chaos or systems with the infinite number particles or the infinite size.

I.e. this Real Dynamics does not depend on errors or exterior noise, and as well as standard Ideal Dynamics depends only on properties of the system.

There are at least two reasons doing Real Dynamics stable against noise: a statistical law of lardge numbers and discreteness of quantum transitions, providing stability of chemical connections [11].

Here there is very important problem how to choose macrovariable. The requirement of this independence of noise also superimposes restriction on a possible choice of macrovariables. For example, except for a pair of states «Schrodinger cat » like alive and dead we can choose their half-sum and a difference in a quantum mechanics. Why in a reality the select{choice} remains only behind a pair a cat alive or dead? Just because this pair is stable against small noise of a macroscopic environment, while their half-sum or a difference break up even at very small exterior noise (theorem Daneri-Loinger-Prosperi [30], [31] Daneri A., Loinger A., Prosperi G. M.) [14].

Other restrictions on macrovariables, are possible also, for example, because of desire to reduce their number or to make their behaviour more determined.

Other important property of Real Dynamics is ambiguity of a choice of most this dynamics and a set of macrovariable which it features. We develop certain new fundamental laws for the given new level of exposition, only nearly leaning from precise Ideal Dynamics which becomes observationally not checked exactly, but the final and precise choice of these laws is defined only by their convenience to us.

It allows to exchange Unpredictable Dynamics with predicted Real Dynamics of macrovariables which is obtained through introduction "coarsening" or "reductions" or some similar methods. It is necessary to underline especially, that in most cases Real Dynamics is not some approach of Ideal Dynamics (as it is frequently interpreted)the difference between them exists, but it} difference between them does not apparent observationally! I.e. the problem what of these dinamics is correct becomes senseless.

From this independence of result of Real Dynamics, we define frequency of measuring (reduction) at decay of a particle or transition from one energy level on another, also we choose an interval time between measurings (reduction) by such way that in a wide neighbourhood of this interval of time these results did not depend on his time precise quantity, i.e. not too large and not too small, preventing thus paradox kettle which never will begin to boil.

In correspondence with two factors of unpredictability there are two types of Real Dynamics:

1) If the observer and medium are included in system of expositions. It superimposes restriction on precise knowledge of initial conditions of a motion and leads to Real Dynamics that does not dependent on them in a wide interval of their values. It is the most popular type Real Dynamics because such Dynamics has "objective" character and does not depends on exterior factors, though, actually, both types of Real Dynamics are defined only by parameters of the system. To this type widely known « New Dynamics », designed by Prigogine concerns [9], [12].
2) If the observer or medium are not included in system of exposition. Uncontrollable interaction of the exterior observer or an environment with featured system superimposes restriction on precise knowledge of equations of motion because of uncontrollable exterior noise leads to Real Dynamics that does not dependent on this noise in a wide interval of its value and a form. It corresponds to widely usedin quantum mechanics "decoherentization" of the quantum systems, interreacting with exterior "lardge" macrosystems.

Let's consider more in detail on « New Dynamics », designed by Prigogine [9], [12]. It differs from other similar methods a successful choice of "coarsening" procedure. The majority of the real isolated systems in classical mechanics are systems with intermixing where a major part of trajectories are exponentially unstable. Their analog in a quantum case are systems of the infinite size or system with the infinite number of particles. These systems are viewed by Prigogine's theory. Out of this consideration board though restricted, but the important class of classical periodic and almost-periodic systems remains and also almost all quantum isolated systems which almost always possess the same property. It would seem, above mentioned quantum and classical systems viewed by Prigogine are in essence various from this point of view. In infinite quantum systems return time is infinite, and in classical chaos systems has though also casual, but finite quantity. However because of inevitable errors of self-measuring in classical systems starting conditions are spread in a small neighbourhood and due to random quantities of return times the complete return of the system viewed as not one point in a phase space, but all its small neighbourhood is possible too only for the infinite time. For one-point systems these casual return times cannot be selfobserved, as the exterior actual observer will always bring in the correcting error.

Function of phase density possesses property of conservation of phase volume of a initial small neighbourhood. As for systems with intermixing the two close trajectories in one direction exponentially disperse it follows from conservation of phase volume, that in the other direction they should converge so quickly. In this direction also it is offered to do "coarsening" . Its peak quantity is determined by a requirement of independence or very weak dependence of macrovariables from "coarsening" quantity , as it is necessary to do in Real Dynamics. This "coarsening" procedure possesses the remarkable property that differs it from others: equations of motion for the coarsened or not coarsened function of phase density remain the equivalent at this procedure in the sense that it is permutation with "coarsening" procedure i.e. is not important what to make in the beginning: to carry out procedure "coarsening" and to use Prigogine equations for deriving the final coarsened function of phase density or to use Ideal Dynamics, and in the end to make this "coarsening".

At the opposite velocities, the square enveloped by not coarsened function of phase density does not change. For usual methods of "coarsening" this property is conserved, only reversibility of equations of motion becomes broken that is not observed in real situations, as already it was many times spoken. In Prigozhin's equations nonreversibility appears because of asymmetry of procedure "coarsening" for opposite velocities. Really, as "coarsening" procedure is carried out in a direction of converging, at the opposite velocities a direction of converging becomes a direction of expansion, i.e. a direction of an expansion of "branches" of function of the phase density, generatored by "intermixing", in this case its square becomes almost unchanged , in contrast with not opposite velocities variant where the direction of "coarsening" is orthogonal and perpendicular to these "branches". The entropy of the coarsened function will decrease with the increase of square enveloped by this function of phase density that leads to to its increase in Prigogine equations, because for the case opposite velocity after "coarsening" the density function has smaller square, as it was considered , that leads to to nonreversibility of the Prigogine equations for the coarsened function of phase density. What is real quantity of "coarsening"? It has value enough to provide an real experimental indistinguishability of "New" Real Dynamics and Ideal Dynamics during the infinite time of observation for the described systems with intermixing. This fact has not been marked by Prigogine's school.

By the way, similar "coarsening" can be made and for periodic or nearly periodic systems, but an indistinguishability Ideal and "New" Dynamics will take place in this case only during time equal to half period of return, that actually it is enough, because It was shown above as even already these returns for real systems are not observational. This fact also has not been noted by Prigogine's school. What quantity "coarsening" ? it is so to provide real experimental indistinguishability of "New" Real Dynamics and Ideal Dynamics during a half-period for described periodic or nearly periodic systems with intermixing during only a half-period.

Many disputes arise in connection with a problem, what is truly ideal Dynamics or Real « New Dynamics » of Prigogine [28]? This dispute is very similar on a dispute what around of what the Earth around of the Sun or on the contrary rotates? Actually, by the definition of a motion the choice remains for us and is determined only by beauty and our convenience, is similar to what in a mathematical science determines a select of definitions and theorems, and the theory of this select, by the way, is still unopened continent of its bases, as against the theorem of Godel! **(the Note about Fermat theorem: By the way, by effect of the theorem of Godel difficulty to prove Fermat theorem was tried to explain , and final result appeared is very close to this statement. Fermat theorem is obtained as a corollary of much more universal and wide theorem, than it [13], and also including on the order is more than postulates, than the theory of natural numbers, on the basis of which Fermat theorem can be formulated. By the way to tell, it would be extremely curious to receive from mathematicians not only 150 page proof, but also the complete and quite visible list of postulates of mathematics, with the indication what from them were used and what are not present for this proof. Similarly to this, the proof of selfconsistency of arithmetics is gained by introduction transfinite induction and corresponding expansion of the list of postulates. To tell the truth, here there is no generalized theorem, a new postulate of transfinite induction is introduced by "hands". What to assume a postulate, and that the theore it is business of beauty and convenience. And the beauty will rescue the world!)**

Similarly to this, the difference between "New" and Ideal Dynamics in the majority of actual situations is unobservable in real experiment so a choice what is truth is arbitrary. For rare cases of "large" systems when Ideal Dynamics can be precisely checked it always meanwhile wins. May be the explored systems yet not large enough? The answer behind the further experiments. That appears valid and what Dynamics from all possible Dynacises will win it is possible to guess only. The situation here is similar to Great Superstring Theories and Grand Unification: now only guesses, and we must wait for experiment which should give the answer may be hundreds years, if only Newcomers or Space will not help us. By the way, in theory of Einstein gravitation, which is considered as doubtless truth (here it, force of authority!), we have while in experiment only hints on true (we shall recollect theories of gravitation of Mordehai Milgrom or Logunov and mysterious space dark matter). How many to wait here? Only the God knows.

Let's finish this part the important note. Together with systems, featured by Real or Ideal Dynamics, systems can exist which, whan we want to describe them explicitly and in details, are featured by Unpredictable Dynamics, i.e. their detailed description is theoretically possible, but is not checked experimentally and so not implemented in practice. Probably alive systems also concern to such type of systems.

**Part 5. A life and death.**

Let's mark from the very beginning, that if the previous parts had more or less strict character, given part by virtue of the obvious reasons has more hypothetical character and is more likely a set of hypotheses.

Let's assume here from a hypothese, that a life completely corresponds to laws of physics.

What is the life and death from the point of view of physics?

Whether there are at an alive substance certain properties not compatible with physics?

Than alive systems differ from lifeless from the point of view of physics?

When alive systems have consciousness and freedom of will from the point of view of physics?

The life is defined, usually, as the special highly organized shape of existence of the organic molecules, possessing ability to metabolism, reproduction, adaptations, motion, response on exterior excitator, ability to self-preservation during long time or even to increase a level of self-organizing. This valid, but too narrow definition: many from alive systems possess only a part from these properties, some of them are proper also in a lifeless substance, inorganic shapes of a life are quite possible also.

The first attempt to describe a life from the point of view of physics has given by Schrodinger [11]. In the book a life has defined as an aperiodic crystal, i.e. high order (and, accordingly, possessing a low entropy and "eating" negentropy from an environment, i.e. in essence unclosed system), but not based on simple regularity, as against a crystal, form of a substance.It also has given there two reasons doin Real Dynamics of alive systems stable in their interior and exterior noise: a statistical law of large numbers and the discreteness of quantum transitions, providing a stability of chemical connections. A principle of activity of alive organisms is similar to clocks: at both cases « an order from the order » appear despite of high temperature.

In the book of the Soviet biophysicist Bauer [12] has determined, that not only high orderliness (and, accordingly, a low entropy) are exhibited not only in a non-equilibrium of allocation of substances in an alive substance, but also the structure of an alive substance is low entropy and strongly unstable. This unstable structure not only is supported due to process of a metabolism, but also is their catalytic agent. This guess correctly only in part, for example proteins or viruses maintain the structure also in the crystalline shape, but them low entropy and strongly unstable modifications and combinations inside an alive substance possess this property. Eventually, nevertheless, there is a gradual degradation of structure, as leads to to inevitability of death and necessity of reproduction for maintenance of a life. I.e. process of a metabolism only very strongly decelerates decay of the composite structure of an alive substance, instead of supports it all time constant. The experimental results given by Bauer, confirm an energy liberation and accordingly increasing entropy in process autolysis, i.e. decay of an alive substance. the reason of this increasing at the first stage of process in instability of the most initial structure without a metabolism supporting it and on the second stage of process in operation protolytic (decomposing) enzymes relieved or appearing at autolysis. Bauer also assume that presence of this redundant structural energy is typical property of a life.

In all these books consideration a separate alive organism is given, while a life as set of all organisms as a whole (biosphere) can be surveyed and defined. Also the problem on an origin and sorce of a life is here concerns. Most the complete and modern answer to these problems from the point of view of physics has been given in Elitzur paper[18]. In this paper he considers a sorce of a life as ensemble of selfreplication molecules. Transiting through a bolter of Darvinian natural selection, a life accumulates in the genes the information (or more likely knowledge in the definition given Elitzur) about a surrounding medium, raising, thus, a level of the organization (negentropy) according to the second law of thermodynamics. lamarckism in its too straight form it is proven in an inconsistency with this law of physics. Wide spectrum of papers and books in this field is discussed in paper. The disadvantages of this paper are follow:

1) Consideration is correct for a life as a whole, as appearances, but not for separately taken alive organism.
2) The suggested proof rejects only too rough, rectilinear model lamarckism while there are many hypotheses and experience, featuring a possibility of embodying of its theory even in an real life [32].
3) For the self-organizing dissipative systems suggested by Prigogine, for example Benard's cells, property of adaptation, as against alive organisms is negated. Naturally, their adaptive abilities are not comparable to alive systems, but in the rudimentary shape, nevertheless, exist. So, for example, Benard's cells, change the geometry or even disappear, as function of an difference of temperatures between the down and upper stratum of a fluid. It also is the primitive shape of adaptation.

The equilibrium ensemble were in equilibrium with a calorstat, in a quantum mechanics in energy representation is featured by a diagonal density matrix. If there are some parameters, except for the energy, systems necessary for the complete description, quantities of diagonal terms to them relevant are equal. Similarly this, in classical mechanics, in equilibrium correlations between the molecules are zero which are analogs of nondiagonal zero elements of a density matrix .

Thus the disbalance is exhibited doubly: in a nonequilibrium distribution of diagonal elements and in an inequality to zero of nondiagonal elements (that corresponds to nonzero correlations in classical mechanics), and these correlations are much more unstable and much faster decay, than a diversion of diagonal devices from equilibrium quantities.

As the life is described by Bauer as self-preserving due to a motion and a metabolism strong instability, we can assume, that the majority of this instability results from these strongly unstable correlations (in a quantum mechanics of nondiagonal elements), which alive systems aspire to support and maintain during time their much greater relaxation time. In lifeless systems it is achieved primely by isolation of system, in the alive unclosed systems actively interreacting with an environment it is achieved by their exterior both interior motion and a metabolism. In it alive systems are similar to isolated systems in which correlations are passively maintained due to this isolation, in alive unclosed systems they are supported due to their active interaction with an environment. It is necessary to note, that alive systems support correlations as inside, and correlations with an environment.

Let's enter concept of pseudo-alive physical systems. We shall term as those simple physical systems describing in the rudimentary shape certain real or guessed properties of alive systems. So, growth of crystals simulates ability of alive systems to replication. By the way, the analysis of these systems allows to find a feeble place in Wigner arguments [6], [27], seeing an inconsistency between ability to replication and a quantum mechanics.

Other example are the quantum isolated systems showing property of maintenance of correlations, similar to maintaining of the strong instability in alive systems, connected to maintenance of correlations or nonzero nondiagonal elements of a density matrix.

To tell the truth, this maintenance is passive. The active deceleration of relaxation times of nondiagonal elements of a density matrix, more the close to methods of maintaining of correlations in alive systems, is achieved in such unclosed systems, such as micromasers [25] which are one more example of pseudo-alive systems. The dissipative systems illustrate properties of unclosed alive systems to maintaining a low entropy and primitive adaptation to a modification of conditions of a surrounding medium.

By the way, definition of a life as the systems promoting maintenance of correlations in a back balance to exterior noise, well explains mysterious silence of SPACE, i.e. absence of signals from other reasonable worlds. The Universe has taken place from uniform centre (Big Bang) and all its parts are correlated, the life only supports these correlations and exists on their basis. Therefore processes of origin of a life in various parts correlated and also are on one level of development, i.e. supercivilizations, capable to reach the Earth while are not present currently. Effects of long-range correlations it is possible to explain and a part indeed wonderful displays of human intuition and parapsychological effects. And here the quantum mechanics is unessential, similar correlations are proper also in the classical mechanics having analogs of nondiagonal elements of a density matrix. Similar correlations frequently wrongly coordinate only with a quantum mechanics.

The following contribution to understanding of a life was made by Bohr [26]. He has paid attention, that the complete measuring of a state of system imports in a quantum mechanics inevitable contortions to behaviour of system, than probably the property of principal nonperception of life can be explained . The criticism of these opinion of Bohr by Schrodinger [19] is not well-grounded. It is based that the complete measuring of a state of system is possible in a quantum mechanics, it is only different from classical mechanics because of its random feature. A problem not that such measuring is impossible. The true problem consists that similar measuring changes the further behaviour of system, in labsence of measuring it would be other [20]. Measuring breaks thin correlations between parts of system, changing its behaviour. It concerns not only to a quantum mechanics, but also to classical mechanics where between real systems there is a finite interaction.

Pseudo-physical systems describing the property of measuring to break dynamics of systems are oscillating the quantum almost isolated systems changing under the scheme: A-> sum of A and B-> B-> difference A and B> A, where A and B - states of system. Measuring in what state is system A or B breaks states of their sum or difference, changing real dynamics of system and destroying correlations (nondiagonal elements of a density matrix) between A and B in these states [10].

Successes of the molecular genetics do not refute this point of view. Build-up of Real Dynamics of a life basically is possible. Really, alive systems are the unclosed systems actively interreacting with a casual environment. The exterior observer interreacts with them usually much more feeblly and cannot cause the key modification in their behaviour. However attempt to understand and predict a life too explicitly in details breaks the thin correlations maintained by a life, and will lead to to Unpredictable Dynamics of alive systems, given the effect predicted by Bohr. Probably, especially thin human intuition and some parapsychological effects also lay in this field of Unpredictability. That they can lay only in this narrow field on the verge of apprehensibility by the exact science and it does not allow to natural selection to strengthen these properties, and does not allow us to understand completely these appearances by help of a science [21], [22].

**Part 6. The Conclusion.**

The given paper is not only philosophical abstract build-up. Not understanding of these bases leads to to errors. As examples made in the theory of poles errors it can serve in problems of a motion of flame front and growth of "finger" on an interface of fluids.

Sivashinsky [23] stated, that Ideal Dynamics of poles leads to acceleration of flame front , and this acceleration is not caused by noise as it does not vary at decreasing of noise and depends only on properties of the system. But in fact also Real Dynamics, connected to noise, does not depend on it in a wide interval of values.

Tanveer [24] has found distinction in growth of "finger" in the theory and numerical experiments, not having understood, that this differences is connected to the numerical noise giving in new Real Dynamics.

It only two ordinary examples, taken of daily practice of the author of paper and it is possible to meet many such examples yet. Results enunciated in this paper are necessary for understanding of base of nonlinear dynamics, thermodynamics and a quantum mechanics.

**Bibliography.**

**Application A. Density matrix**

Let's consider a beam of $N_a$ particles prepared in a state $|\chi_a\rangle$ , and yet independent from the first beam the second beam of $N_b$ the particles prepared in a state $|\chi_b\rangle$ . For description of an integrated beam we shall enter the operator ρ of mixed state, defined by expression

$$\rho=W_a|\chi_a\rangle\langle\chi_b| + W_b|\chi_a\rangle\langle\chi_b|$$

Where $W_a=N_a/N$, $W_b=N_b/N$, $N=N_a+N_b$. The operator ρ term as the operator of density or the statistical operator. It features method of preparation of beams and so contains the information on the complete beam. In this sense the intermixture is completely described by the operator of density. In that specific case a pure state $|\chi\rangle$ the operator of density is given by expression

$$\rho = |\chi\rangle\langle\chi|.$$

It is more convenient usually to write the operator ρ in the matrix shape. For this purpose we shall choose set of basis states (usually $|+1/2\rangle$ and $|-1/2\rangle$) and it is decomposable states $|\chi_a\rangle$ and $|\chi_b\rangle$ on this set according to a relation

$$|\chi_a\rangle = a_1^{(a)}|+1/2\rangle + a_2^{(a)}|-1/2\rangle,$$
$$|\chi_b\rangle = a_1^{(b)}|+1/2\rangle + a_2^{(b)}|-1/2\rangle.$$

In representation of states $|\pm 1/2\rangle$:

$$|\chi_a\rangle = \begin{pmatrix} a_1^{(a)} \\ a_2^{(a)} \end{pmatrix}$$

$$|\chi_b\rangle = \begin{pmatrix} a_1^{(b)} \\ a_2^{(b)} \end{pmatrix}$$

And for the conjugate states-

$$\langle\chi_a| = (a_1^{(a)*},\ a_2^{(a)*}),$$
$$\langle\chi_b| = (a_1^{(b)*},\ a_2^{(b)*}).$$

Applying rules of multiplication of matrixes, we shall receive for "outer product"

$$|\chi_a\rangle\langle\chi_a| = \begin{pmatrix} a_1^{(a)} \\ a_2^{(a)} \end{pmatrix} (a_1^{(a)*},\ a_2^{(a)*}) = \begin{pmatrix} |a_1^{(a)}|^2 & a_1^{(a)} a_2^{(a)*} \\ a_1^{(a)*} a_2^{(a)} & |a_2^{(a)}|^2 \end{pmatrix}$$

And similar expression for product $|\chi_b\rangle\langle\chi_b|$ . Substituting these expressions in the functional of density, we discover
Density matrix

$$\rho = \begin{pmatrix} W_a \left|a_1^{(a)}\right|^2 + W_b \left|a_1^{(b)}\right|^2 & W_a a_1^{(a)} a_2^{(a)*} + W_b a_1^{(b)} a_2^{(b)*} \\ W_a a_1^{(a)*} a_2^{(a)} + W_b a_1^{(b)*} a_2^{(b)} & W_a \left|a_2^{(a)}\right|^2 + W_b \left|a_2^{(b)}\right|^2 \end{pmatrix}$$

As at a deduction of this expression basis states $|\pm 1/2\rangle$ were used, the obtained expression term as a density matrix in $\{|\pm 1/2\rangle\}$ representation.

In summary some words about statistical matrix $P_0$, possessing remarkable properties. We know, that in classical statistical thermodynamics all possible macroscopic states of system are considered as a priori equality probability (another words, they are considered equality probability if there are no no informations about value of a total energy or about contact to a calorstat supporting a stationary value temperature of system, etc.). By analogy to it in a wave mechanics all states of system, defined the various functions, generator a complete set of orthonormal functions, it is possible to guess a priori equality probability. Let $\varphi_1,\ldots,\varphi_k$, - such system of basis functions $\varphi_k$; knowing, that the system is characterized by an intermixture of states $\varphi_k$, in absence of any other information it is possible to think, that the statistical matrix of system looks like

$P_0=\sum_k pP_{\varphi_k}$ , где $\sum_k p = 1$ ,

I.e., that $P_0$ - a statistical matrix of such mixed state for which all of a weight are equal among themselves. We accept $\varphi_k$ for basis functions, matrix $P_0$ can be presented as

$(P_0)_{kl}=p\delta_{kl}$

If the statistical state of ensemble of systems in an initial instant is characterized by matrix $P_0$ and if in all systems of ensemble to carry out measuring same quantity A the statistical state of ensemble as former will be characterized by matrix $P_0$.
Equations of motion for a density matrix ρ

$$i\frac{\partial \rho_N}{\partial t} = L\rho_N$$

Where L - the linear functional:
$L\rho = H\rho - \rho H = [H,\rho]$,
Where H - the functional of energy of system.

If A - operator of some observer value,

mean value of this value can be found as follow:
$\langle A\rangle = trA\rho$

**Application B**. **Reduction of a density matrix and the theory of measuring.**
Let at measuring some object we « precisely distinguish » states $\sigma^{(1)}$, $\sigma^{(2)}$, ... . Yielding measurings above object were in these states, we gain numbers $\lambda_1$, $\lambda_2$, ... . An initial state of the measuring device let's designate through a. If the measured system at the beginning was in a state $\sigma^{(\nu)}$ a state of a complete set « object of measuring plus the measuring device » before they have entered interaction will be determined by direct product a X $\sigma^{(\nu)}$, . After measuring:

$$a \times \sigma^{(\nu)} \rightarrow a^{(\nu)} \times \sigma^{(\nu)}$$

Let now the initial state will be not "precisely distinctive", and the arbitrary combination $\alpha_1 \sigma^{(1)} + \alpha_2 \sigma^{(2)}+$... such states. In this case from linearity of the quantum equations follows

$$a \times [\Sigma\alpha_\nu\sigma^{(\nu)}] \rightarrow \Sigma\alpha_\nu[a^{(\nu)} \times \sigma^{(\nu)}]$$

In the final state incipient as a result of measuring there is a statistical correlation between a state of object and a state of the device: simultaneous measuring at system « object of measuring plus the measuring device » two quantities (the first is measured investigated object and the second is place of device arrow) always leads to compounded results. Thereof one of termed measurings becomes excessive - about a state of object of measuring it is possible to obtain on the basis of observation over the device.
Result of measuring is not the vector of a state represented as sum, being in a right part of the found relation, and a so-called intermixture, i.e. one of vectors of a state of type

$$a^{(\nu)} \times \sigma^{(\nu)}$$

Also that at interaction between object of measuring and the measuring device this state arises with probability $|\alpha_\nu|^2$. The given transition is termed as a reduction of a wave

packet and corresponds to transition of a density matrix from nondiagonal $\alpha_\nu\alpha_\mu^*$ in a diagonal form $|\alpha_\nu|^2\delta_{\nu\mu}$. This transition is not featured by quantum equations of motion.

**Appendix C Function of phase density**

At present time t the state of system N of identical particles from the point of view of classical mechanics is defined{determined} by the representation of values of coordinates $\mathbf{r}_1, \ldots, \mathbf{r}_N$ and impulses $\mathbf{p}_1, \ldots, \mathbf{p}_N$ all N particles of system. We shall use for brevity a label

$$x_i=(\mathbf{r}_i,\mathbf{p}_i) \quad (i=1, 2, \ldots, N)$$

for set of values of coordinates and components of impulse of a separate particle and a label

$$X= (x_1, \ldots, x_N)=(\mathbf{r}_1, \ldots, \mathbf{r}_N, \mathbf{p}_1, \ldots, \mathbf{p}_N)$$

for set of values of coordinates and values of impulses of all particles of system. Such space 6N variable term as a 6N-dimensional phase space.

To define concept of a distribution function of states of system, we shall consider a set of identical macroscopic systems - ensemble of Gibbs. To all these systems of a requirement of experiment are identical. As, however, these requirements define a state of system ambiguously to different states of ensemble at present to time t there will correspond different values X.

Let's allocate in a phase space volume dX about point X. Let at present time t in this volume the points describing states dM of systems of ensemble from their complete number M are made. Then a limit of the ration of these quantities

$$\lim_{m\to\infty} dM/M=f_N(X,t)dX$$

defines a phase density function in an moment t.

$$\int f_N(X,t)dX=1$$

The equation of Liouville for function of phase density it is possible to note as

$$i\frac{\partial f_N}{\partial t} = Lf_N = \{H, f_N\}$$

Where L - the linear functional:

$$L = -i\frac{\partial H}{\partial p}\frac{\partial}{\partial x} + i\frac{\partial H}{\partial x}\frac{\partial}{\partial p},$$

Where H - energy of system.

**Application D "Coarsening" function of phase density and a hypothesis of the molecular chaos. "Coarsening"** of a density function we shall term its replacement on approximate quantity.

For example

$$f_N^*(X,t)=\int_{(y)}g(X-Y)f_N(Y,t)$$

Where

$$g(X) = 1/\Delta\, D(X/\Delta)$$

$$D(x) = 1 \text{ for } |X| < 1$$

$$D(x) = 0 \text{ for } |X| \geq 1$$

Other example "coarsening" - « a hypothesis of the molecular chaos »

Replacement of a two-particle distribution function on product of one-particle functions
$f(x_1, x_2, t) \to f(x_1,t) f(x_2,t)$

**Application E** of Definition of an entropy
As definition of an entropy we shall specify expression

$S = -k \int_{(X)} f_N(X, t) \ln f_N (X, t)$

For a quantum mechanics through a density matrix

$S = -k \, tr \, \rho \ln \rho$ [29]

where tr - a track of a matrix.

These entropies does not change under reversibal evolution.
To obtain changing values we must use instead functions $f_N$ or ρ their "coarsened" functions.
.

**Application F New Prigogine's dynamics.**
Let to enter the linear functional Λ on function ρ phase density or a density matrix
$\tilde{\rho} = \Lambda^{-1} \rho$
$\Lambda^{-1} 1 = 1$
$\int \tilde{\rho} = \int \rho$
$\Lambda^{-1}$ maintains a positiveness.
Such, that function $\tilde{\rho}$ possesses property: for functions Ω particular as
$\Omega = tr \tilde{\rho}^{+} \tilde{\rho}$ or $\Omega = tr \tilde{\rho} \ln \tilde{\rho}$
We have $d\Omega/dt \leq 0$
Equation of motion for ρ

$$\frac{\partial \tilde{\rho}}{\partial t} = \Phi \tilde{\rho}$$

Where $\Phi = \Lambda^{-1} L \Lambda$
Φ- a nonreversible Markov semigroup
$\Lambda^{-1}(L) = \Lambda^{+}(-L)$
For function of phase density the functional Λ-1 corresponds "coarsened" in a direction of squeezing of phase volume. For a quantum mechanics such functional can be found only for the infinite volume or the infinite number of particles. We shall enter in a quantum mechanics projecting functional P zeroing nondiagonal devices of a density matrix. The functional Φ and basis vectors of a density matrix are choosed so that to make the functional Φ permutable with functional P:

$$\frac{\partial \mathrm{P}\tilde{\rho}}{\partial t} = \mathrm{P}\Phi\tilde{\rho} = \Phi\mathrm{P}\tilde{\rho}$$

**Application G Impossibility of introspection and self-calculation.**
Let's assume, that there is a powerful computer capable to predict the future and its environments on the basis of calculation of a motion of all molecules. Let its prediction is rolling up a black or white ball from some devices. The device rolls out a white ball if the machine predicts black and on the contrary a black ball at a prediction white. Clearly, that a prediction of the machine always are not true, that proves impossibility of introspection and self-calculation.

**Application H Correspondence of a quantum and classical mechanics. The table.**

| Quantum mechanics | Classical mechanics |
|---|---|
| Density matrix | Function of phase density |
| Equation of motion for density matrixes | the Equation of Liouville |
| Reduction of a wave packet | "Coarsening" functions phase densities or a hypothesis of molecular chaos |
| Inherent action of the observer or environments, featured by a reduction | it is theoretically indefinite small, but in a reality finite interaction of systems with the observer or an environment |
| Correlations between velocities and positions of molecules in different parts of system | nonzero nondiagonal elements of a density matrix |

# Основы нелинейной динамики или Реальная Динамика, Идеальная Динамика, Непредсказуемая Динамика и “Шредингеровский кот”.

**Абстракт.**
В статье обсуждены парадоксы лежащие в основе термодинамики и квантовой механики. Дано их разрешение с точки зрения влияния внешнего наблюдателя (окружающей среды), разрушающего корреляции системы, или ограниченности самопознания системы в случае, когда и наблюдатель, и окружающая среда включены в систему. Введены понятия Реальная Динамика, Идеальная Динамика и Непредсказуемая Динамика. Дано рассмотрение явления жизни с точки зрения этих Динамик.

## Введение.

Статистическая механика и квантовая механика - две разработанные и хорошо известные теории, основа современной физики. Тем не менее они содержат ряд парадоксов, которые заставляют многих ученых усомнится во внутренней непротиворечивости этих теорий.
Перечислим их:

1) Базисные уравнения классической и квантовой механики обратимы во времени в то время как законы термодинамики, например, закон возрастания энтропии в замкнутых системах, необратим, хотя он, по сути дела, должен выводиться из этих уравнений [1], [2].
2) Тоже относится к противоречию между законом возрастания энтропии в замкнутых системах и законом Пуанкаре о возвратах замкнутой системы сколь угодно близко к начальному состоянию [1], [2].
3) Парадокс “Шредингеровского кота” , т.е. редукция волнового пакета и переход из чистого в смешанное состояние в макроскопических системах или в процессе измерения квантовой системы. Этот процесс не описывается уравнениями квантовой механики, совершенно непонятно в какие моменты времени он происходит и какова его длительность.Тем не менее, без него квантовая механика неполна [3].

4) Микроскопическая энтропия классических систем,определенная через функцию плотности в фазовом пространстве, остается неизменной.(Аналогично энтропии квантовых систем, определенной через матрицу плотности [15]) Макроскопическая энтропия систем, определенная через макроскопические переменные или, что математически эквивалентно, «огрубленную» для классической механики функцию плотности в фазовом пространстве ( и редуцированную матрицу плотности для квантовой механики), может как возрастать, так и убывать с одинаковой вероятностью из-за обратимости движения.(В отличие от очень распространенного заблуждения, что процессы с убыванием энтропии менее вероятны. На самом деле, наиболее вероятны состояния с неизменной и максимальной для данной системы энтропией, тогда как состояния, ведущие как к убыванию, так и возрастанию энтропии одинаково редки по сравнению с этими равновесными состояниями) Почему, в таком случае, в реальном мире во всех макроскопических классических системах всегда наблюдается лишь возрастание энтропии? Почему, чтобы получить законы для этого возрастания энтропии нам приходится вводить некие дополнительные предположения ( вроде, например, гипотезы молекулярного хаоса Больцмана или эквивалентного ему по смыслу огрубления функции плотности в фазовом пространстве), помимо основных уравнений классической механики, хотя они и дают нам ее полное описание? [1], [2]

5)Для описания макроскопического состояния используются те или иные макроскопические переменные. Теоретически их выбор ничем не ограничен, это может быть любая функция микроскопических параметров, тем не менее, как правило, выбор "удобных" переменных
достаточно предопределен. Чем?

6)Существует множество живых систем, в том числе обладающих сознанием и свободой воли. Какие физические свойства принципиально отличают живые (сознающие) системы от неживых? Насколько полно они могут быть описаны известными уравнениями физики и термодинамики?

Не даром Шредингер использовал для формулировки своего парадокса
в качестве макроскопического детектора именно живого кота! Некая связь между описанными выше базисными парадоксами физики и тайнами жизни интуитивно ясна многим физикам.Стоит попытаться
сформулировать ее более ясно.

Разрешение этих парадоксов не требует изобретения новых законов физики, хотя их экспериментальная проверка и может привести к ним.
Мы будем анализировать эти проблемы строго в рамках известных законов физики.

Имеется обширная литература посвященная этим парадоксам [1], [2], [3]. Многое из сделанного верно. Проблема в том, что эти рассмотрения не дают ключевого свойства динамической системы, приводящего ко всем этим парадоксам и дают лишь фрагменты их полного решения.

Ситуация тут напоминает парадокс Гиббса [4], где скачок энтропии при смеси газов объясняли отсутствием переходных форм между одинаковыми и разными газами. Тем не менее такие переходные формы возможны, величину скачка можно регулировать и рассмотрение ( не поверхностное) дает не только глубокое понимание и разрешение этого парадокса, но и более глубокое понимание самих основ физики и не требует введения новых законов.

Аналогичная ситуация складывалась и со спином электрона: учебники переполнены утверждениями, что его невозможно интерпретировать как собственный момент именно физического (т.е. в пространстве) вращения частицы. Забавно, тем не менее, что в рамках уравнения Дирака это прекрасно делается и спин может быть истолкован как собственный момент чисто физического вращения волновой функции [5].

Также и вышеуказанные парадоксы могут быть при аккуратном рассмотрении глубоко поняты и дать богатый материал для понимания основ существующей физики с присущими ей ограничениями для объяснения основ окружающего мира. Физика не столь всесильна даже чисто теоретически (как почти все, даже профессиональные физики, уверены) и эти ее ограничения вполне можно вполне вывести из нее самой, не изобретая новых теорий. Эти ограничения и ведут к указанным выше парадоксам. Что забавно, при всем их разнообразии и внешней несвязанности все они имеют единый корень и причину!

Инструментом для нашего рассмотрения будет Нелинейная Динамика, наука которая ставит себе целью НЕ изобретение новых фундаментальных законов физики, а используется для более глубокого понимания уже существующих и для нахождения общих свойств и законов относящихся к совершенно казалось бы разным физическим системам (в смысле предметной области их приложения), но, тем не менее, имеющим очень похожую динамику.

**Часть 1. Квантовый парадокс «Шредингеровского кота».**

Не будем здесь подробно описывать условия этого парадокса, они хорошо описаны во многих источниках [14], [3], [6], [7]. Остановимся здесь на его сути. Кот - это макроскопическая система, которая, будучи описана микроскопически (квантово), может находится одновременно в двух состояниях - кот живой и кот мертвый, являясь их «взвешенной суммой» (т.н. волновым пакетом пси-функций)! В реалии, как возможный сторонний наблюдатель, так и сам кот своим сознанием фиксируют всегда лишь одно из этих двух состояний с вероятностью, определяемой квадратами их «весов» в вышеописанной «взвешенной сумме». На языке математики это соответствует переходу от чистого к смешанному состоянию и диагонализации матрицы плотности.

Этот процесс носит название редукции волнового пакета и не описывается уравнениями квантовой механики. Более того, он соответствует возрастанию микроскопической энтропии (определенной с помощью матрицы плотности), увязываясь тем самым с предыдущими двумя парадоксами и вступая в противоречие с обратимостью квантовых уравнений и "возвратами" квантовой системы. Поскольку эти парадоксы присущи и классической механике это позволит нам найти в дальнейшем также и классический аналог для парадокса «шредингеровского кота», который прекрасно существует, несмотря на всеобщую полную убежденность в его "чисто квантовой природе", связанную с отсутствием глубокого понимания сути и причин этих парадоксов. Необходимость введения этой загадочной редукции волнового пакета для макроскопических систем (являющимися, как правило, и конечными объектами в измерительных системах, т.е. их детекторами) и является основой и сутью парадокса «шредингеровского кота».

Кстати, в литературе встречаются попытки описать средствами искусства некие состояния сознания, отвечающие квантовой «взвешенной сумме» [8]. Но если таковые и существуют на самом деле, они являются скорее экзотикой, чем правилом.

Для разрешения парадокса, казалось бы, проще всего было бы предположить, что эволюция просто распадается на две возможные «ветви эволюции» с определенной вероятностью каждая, аналогично тому, как это происходит в статистической механике, где состояние системы описывается не точкой в фазовом пространстве, а является неким «облаком» таких точек. Дело в том, что редукция волнового пакета, в отличие от «разделения на ветви» эволюции в статистической механике, приводит к иной эволюции ветвей, чем она была бы, сохрани мы «взвешенную сумму». Т.е. эти две возможные ветви эволюции оказывают взаимное влияние друг на друга через смешение результатов итогового измерения (которое по сути и является редукцией), проводимого для «взвешенной суммы» этих ветвей, поскольку решение является арифметической суммой, а не статистической смесью этих ветвей.

Это находит подтверждение в следующих примерах. Первый пример связан с обратимостью квантовых уравнений движения. Как уже было сказано, редукция волнового пакета приводит к возрастанию энтропии, в то время как обратимая квантовая эволюция оставляет ее неизменной. Вряд ли можно рассчитывать на то, что эти два, столь разные, типа эволюции дадут одинаковый результат.

Второй пример связан с теоремой Пуанкаре о «возвратах» динамической системы [1], [2]. Хотелось бы здесь остановиться на некоторых особенностях этой теоремы для квантовых систем. В случае классической механики большинство динамических систем систем являются хаотическими: времена возврата образуют случайную последовательность и сильно меняются по величине при малейших изменениях начальных условий. Тем не менее существует хотя и малый, но физически значимый класс систем с периодическими или почти периодическими временами возврата, устойчивых к малым погрешностям начальных условий. Эти системы интегрируемы в переменных действие-угол.

В случае квантовых систем ситуация прямо обратная - замкнутая система с ограниченным объемом и с конечным числом частиц - всегда обладает периодическими или почти периодическими временами возврата, кроме систем бесконечного объема или бесконечного числа частиц, где это время возврата равно бесконечности [9].

Соответственно, возвращаясь к нашим уравнениям, редукция кота приводит к возрастанию микроскопической энтропии, делая невозможным возврат, поскольку он соответствует прежней, меньшей энтропии, а обычная обратимая квантовая эволюция оставляет энтропию неизменной. Таким образом, редуцированная и не редуцированная системы имеют различную динамику.

Третий забавный пример - «котелок, который никогда не закипит» [3]. Пусть мы имеем частицу, которая должна перейти с верхнего на нижний уровень энергии в соответствие с законами квантовой механики (например распасться). Если проводить акты наблюдения за ней слишком часто она никогда этого не сделает! Искажение эволюции системы редукцией, происходящей в моменты наблюдения, приводит к этому эффекту. Как часто происходят эти эти акты редукции в реальности и какова их собственная временная продолжительность? Не ясно. Кстати, этот парадокс объясняет, почему распад частиц или состояний в квантовой механике не является точно экспоненциальным, а лишь близким к нему. Эти отклонения, казалось бы, позволяют определить «возраст» системы, поскольку лишь экспоненциальный тип распада делает это невозможным. Тем не менее это не так - сам процесс наблюдения и вносимые им искажения из-за редукции делают разницу между экспоненциальным и квантовым типом распада ненаблюдаемой.

Эти рассуждения доказывают, что редукция - это не просто механическое разделение движения на две возможные независимые ветви, она меняет всю реальную динамику системы.

Рассмотрим возможные известные подходы к решению парадокса кота.Все эти подходы представляют собой некие правильные шаги в нужном направлении, но не приводят к полному разрешению парадокса.

Последнее время появилось огромное количество различных интерпретаций квантовой механики, самая известная и популярная из которых «многомировая» [3], которая предполагает, что в момент редукции волнового пакета не выбирается лишь одна ветвь, а они существуют одновременно в неких параллельных мирах, в каждом из этих миров наблюдатель видит лишь одну из ветвей, тем не менее, поскольку они существуют одновременно, когда проводится измерение, все возможные «миры» оказывают влияние на результаты этого измерения, обеспечивая взаимное влияние ветвей, вытекающее из законов квантовой механики, которое при стандартной редукции уничтожается выбором лишь одной из ветвей. Проблема этой интерпретации (впрочем как и всех иных) в том, что она не разрешает сами трудности, а лишь переносит их в иную, менее заметную для нас, плоскость. Так, наблюдатель в любом из возможных «миров» имеет полную информацию лишь о нем самом и ничего в явной форме не знает о других мирах, что делает фактически его информацию неполной и не достаточной для предсказания результатов измерения, которые определяются итоговой эволюцией во всех «мирах». Таким образом, нарушение нормальной квантовой эволюции в случае редукции заменяется принципиальной непредсказуемостью результатов измерения из-за ограниченной информации в каждом из «миров». Хотя намек на правильное разрешение парадокса кота тут есть.

Следующим подходом к разрешению парадокса кота является учет влияния внешнего наблюдателя. Ведь редукция волнового пакета и происходит в момент наблюдения или измерения.Действительно, редукция и нарушение нормального хода квантовой эволюции может быть объяснено в этом случае незамкнутостью системы и воздействием внешних сил.

Тут нам следует остановится на одном из главных принципиальных моменте, отличающем классические и квантовые системы. В классических системах влияние измерительного прибора на исследуемый процесс может быть (конечно чисто теоретически, а не практически) сделано сколь угодно малым. В квантовой механике, измерение не может быть проведено без редукции волновой функции и хоть и малого, но вполне конечного воздействия на измеряемую систему.Чем точнее мы хотим измерить состояние квантовой системы, тем более сильное возмущение мы в нее вносим. Это делает квантовую механику принципиально не замкнутой.Именно в этом, а не в ином типе базисных уравнений или вероятностном характере ее законов, как принято обычно считать, и состоит принципиальное отличие квантовой и классической механик. Следует особо отметить, что это отнюдь не делает невозможным точное обратимое квантовое описание систем или их экспериментальную проверку внешним наблюдателем.Просто это возможно в ограниченной мере, лишь при специфических условиях, которых мы коснемся ниже. Здесь лишь отметим, что возможна проверка внешним наблюдателем лишь правильности самих обратимых квантовых уравнений движения, но не само точное изменение параметров и истории эволюции системы, ими описываемых, поскольку они определяются не только видом уравнений, но и начальным состоянием системы, неизбежно нарушаемым наблюдением.

Невозможность исключить влияние измерения иллюстрируется опытом с квантом, падающим на экран с двумя щелями, и интерференционной картиной за ними: при измерении через какую из двух щелей прошел квант влияние второй щели становится невозможным исходя из максимальности скорости света и принципа причинности, и интерференционная картина исчезает, заменяясь простой суммой интенсивностей сигналов от двух щелей. Для придания процессу взаимодействия с наблюдателем объективности и независимости от субъективного наблюдателя его воздействие может быть заменено неконтролируемым взаимодействием с макроскопической окружающей средой, что также вносит элемент незамкнутости в систему. Этот процесс носит название декогеренизации [16] и пользуется заслуженной популярностью. Его главное преимущество состоит в том, что он заменяет набор квантовых обратимых уравнений движения для замкнутой системы и таинственный необратимый процесс редукции замкнутой системой уравнений, включающей внешний шум. Эти уравнения имеют конкретное и чисто практическое значение. Здесь-то нас и поджидают новые трудности. Менее принципиальная состоит в том, что параметры системы и уравнения движения становятся зависимыми от конкретного вида некого неконтролируемого внешнего воздействия. Каким оно должно быть выбрано, чтобы обеспечить верное и правильное описание системы?

Гораздо более принципиальная трудность состоит в том, что как наблюдатель, так и внешняя среда могут быть в принципе включены в саму систему и процесс редукции, который реально наблюдается и никуда не исчезает и в этом случае, будет происходить уже в замкнутой системе.В этом случае, найденное выше объяснение редукции, как некого внешнего воздействия, отпадает и мы возвращаемся к нашему парадоксу в его прежней нерешенной форме. Простейшим примером для понимания этого явления является случай, когда сам кот занимается самонаблюдением, констатируя, что он сам еще жив, а не мертв.В этом случае конечным этапом измерения будет сознание кота. Что же делать? Вводить бесконечную последовательную цепочку наблюдателей, расширяя нашу систему до бесконечности? Приписать нашему сознанию реальную материальную силу и способность к редукции системы или даже, как иногда предлагается, создать некую новую физику сознания [7]? Все гораздо проще и красивее.

В случае, когда кот находится под внешним наблюдением и наблюдатель или среда не включены в систему, как мы видели, парадокса не возникает. Проблема тут состоит в том, чтобы четко определить конкретный возможный вид внешнего воздействия, что точно невозможно, поскольку воздействующие объекты не включены в рассмотрение. Что делать когда все внешние факторы входят в систему, в том числе, когда наблюдателем будет сам кот? Эволюция и динамика системы будет четко различна для случая обратимой динамики без редукции и необратимой с редукцией. Наиболее ярко это иллюстрируется отсутствием возврата системы в исходное состояние при наличие редукции из-за возрастания энтропии в этом процессе и возвратами, причем почти периодическими, предсказываемыми обратимыми уравнениями квантовой механики. Разрешение этого парадокса заключается в том, что эта разница между двумя динамиками хоть реально и существует,но не может быть проверена на практике, т.е. экспериментально.Действительно, поскольку замкнутая изолированная система не может ни точно и однозначно померить свое полное состояние, ни решить систему уравнений описывающих их динамику, она не способна также ни полностью экспериментально проверить собственную динамику, ни сделать однозначный выбор между двумя возможными динамиками, как в данном конкретном случае, поскольку в проверяемой области они дают идентичный результат. Так, система не может проверить свой собственный возврат в начальное состояние, поскольку, чтобы сделать эту проверку, она должна помнить, что прежде уже находилась в этом состоянии некоторое время назад, но подобная «память» должна иметь некий материальный носитель, который никак не мог существовать в начальном состоянии, в которое система, как мы полагаем, вернулась, поскольку в этом случае и сама «память» возвращается в прежнее состояние, таким образом «забывая» всю «историю». Т.е., вернувшись в прежнее состояние, система должна неизбежно потерять память о всей своей истории и не сможет зарегистрировать возврат. Только внешний независимый наблюдатель с независимой внешней памятью способен на подобную проверку. Он же (или внешняя среда вместо него) и внесет неизбежные в квантовой механике погрешности в измеряемую систему, которые и объяснят редукцию. Таким образом, наличие двух различных динамик для одной и той же системы объясняется невозможность чисто экспериментально обнаружить разницу между ними, а не загадочными созидательными силами сознания.

Здесь важно отметить, что основой этого парадокса и причиной, что он столь явно проявился именно в квантовой механике, является то, что как раз в квантовой механике измерение неизбежно приводит, пускай, как правило, к малому , но конечному возмущению измеряемой системы. Именно наличие подобного возмущения, а не ее вероятностный характер, и отличает принципиально квантовую механику от классической, приводя к очень трудноразрешимым парадоксам.

Здесь стоит еще раз отметить, что, тем не менее, при строго определенных условиях, которые будут описаны ниже, точная проверка обратимых квантовых законов возможна, несмотря на это ограничение.

## Часть 2. Классический аналог квантового парадокса «Шредингеровского кота».

Парадокс «Шредингеровского кота», как правило, рассматривается, как чисто квантовый, не имеющий аналога в классической механике. Это, тем не менее, заблуждение. Аналог редукция волнового пакета имеет место и в классической механике! Действительно, закон возрастание энтропии ( в квантовой механике происходящий в момент редукция волнового пакета) происходит и в классической статистической механике, и по аналогии с квантовой механикой, входит в противоречие с обратимым характером законов движения и теоремой Пуанкаре о возвратах.Эти классические законы движения оставляют энтропию неизменной (если она определена через функцию плотности в фазовом пространстве), или она может как возрастать, так и убывать (если она определена через «огрубленную» функцию плотности в фазовом пространстве, т.е. усредненную в некоторой окрестности каждой точки функции плотности в фазовом пространстве). Чем же объясняется только возрастание энтропии в реальности? Введением Больцмановской «гипотезы молекулярного хаоса» [1], [2], которая по сути и является одним из видов «огрубления» функции плотности в фазовом пространстве, приводящему к потере корреляций между скоростями и положением разных молекул ( что, по сути, выражается в обратимости движения при обращении скоростей и возвратах через известное время в малую окрестность начального состояния) и необратимой эволюции системы. Но мы здесь видим полную аналогию с редукцией волнового пакета! Она также приводит к потере корреляций (недиагональных элементов матрицы плотности) и необратимости движения! Т.е. «гипотеза молекулярного хаоса» или иной тип «огрубления» функции плотности в фазовом пространстве и является полным аналогом редукции в квантовой механике. Потеря корреляций при этих операциях эквивалентна обнулению недиагональных элементов матрицы плотности при редукции волнового пакета. Чем же можно объяснить введение этих дополнительных к классической механике и противоречащих им огрублений»? Да теми же причинами, что и редукция в квантовой механике: неразличимостью двух видов динамик(с и без огрублением) при реальных экспериментах (для замкнутых систем из-за невозможности «запоминания» начальных условий при самонаблюдении из-за возвратов, для внешнего же наблюдения взаимодействием наблюдателя или окружения с наблюдаемой системой) . Но тут мы и сталкиваемся с главным и принципиальным различием классической и квантовой механик.

Если в квантовой механике взаимодействие наблюдателя с наблюдаемой системой всегда присутствует и конечно, то в классической механике оно может быть теоретически сведено к нулю! На деле же оно всегда имеет место и конечно. Этим и объясняется противоречие между теоретической наблюдаемостью убывания энтропии и ее отсутствием в реальности в больших системах: реальное, конечное и малое взаимодействие с наблюдателем или просто «окружением» приводит к разрушению процессов с убыванием энтропии. Действительно, процессы с убыванием энтропии, в отличие от процессов с возрастанием энтропии,сильно неустойчивы по отношению к хаотическому внешнему воздействию, что приводит к их разрушению и синхронизации стрел времени между наблюдателем, наблюдаемой системой и их окружением, даже для малого взаимодействия. Стрела времени определяется в направлении возрастания энтропии. Введением этих стрел времени и пытались объяснить ранее закон возрастания энтропии. При этом появлялся законный вопрос: поскольку оба направления стрелы времени равновероятны (максимально вероятны лишь равновесные состояния без явно выраженного направления) почему в реальности все эти стрелы сонаправлены? Это считалось неразрешимой загадкой, решение которой необходимо искать чуть ли не глубинах происхождения Вселенной [28], в то время как решение очень просто и лежит лишь в реальном, пусть и малом взаимодействии всех подсистем, приводящем к всеобщей синхронизации всех стрел времени**.** Надо отметить, что теоретическая возможность нулевого взаимодействия между системами в классической механике и привела к тому, что все эти явления, включая парадокс «Шредингеровского кота» и проявились в явной и четкой форме лишь в рамках квантовой механики, в то время как они характерны и для классической механики при постулировании малого, но конечного взаимодействия между системами, которое всегда имеет место в реальном мире, за исключением некоторых очень тонких и искусственных ситуаций, создаваемых самими людьми в экспериментах. К этим ситуациям мы сейчас и перейдем. Предварительно лишь условимся, что в дальнейшем тексте называя системы «реальными», мы будем подразумевать наличие такого взаимодействия, хоть и малого, с окружением или наблюдателем, хотя в классической механике оно может быть и нулевым.

**Часть 3. Экспериментальное наблюдение квантового парадокса «Шредингеровского кота» и экспериментальная проверка Идеальной динамики.**

В классической механике такая ситуация создается очень просто: почти идеальной изоляцией системы от внешнего окружения и введением почти нулевого взаимодействия системы с наблюдателем. В квантовой механике это невозможно: измерение системы всегда связанно с неустранимым взаимодействием. Может показаться, что разница между редуцированной или не редуцированной динамикой ( будем не редуцированную динамику (в классической механике динамика до введения «огрубления») в дальнейшем называть Идеальной динамикой) становится в любом случае реально не проверяемой. Этот вывод, тем не менее, ошибочен. Измерение в квантовой механике возможно двух видов: когда результат измерения соответствует состоянию системы до или после измерения с теоретически 100% точностью. Оба состояния одновременно измерены быть не могут из-за взаимодействия с наблюдателем. Причем измерение, соответствующее состоянию системы после измерения является скорее не измерением, а «подготовкой» системы к измерению, поскольку исходное состояние меняется и остается неизвестным. Измерение, соответствующее состоянию системы после измерения будем по сути именовать «наблюдением». Таким образом, квантовая система с «подготовкой» в начальный момент, почти полной изоляцией в промежуточном состоянии от наблюдения и внешней среды и «наблюдением» в конечном состоянии и является по сути полигоном для проверки Идеальной динамики, квантовым эквивалентом классической изолированой системы.
Она имеет следующие существенные недостатки:

1) исходное состояние меняется и остается неизвестным.
2) промежуточные состояния не измеряются и остаются неизвестны.По сути можно сравнить на соответствие Идеальной динамике лишь начальное и конечное состояния.
3) случаи, когда можно добиться столь полной изоляции редки и требуют огромных усилий.

Примеры таких систем:

1) Мезоскопические системы при низких температурах [10].

Эти системы за счет больших размеров близки к границе применимости закона больших чисел и почти макроскопичны. За счет низких температур(и, соответственно, импульсов и скоростей молекул) размеры квантового волнового пакета велики (в силу соотношения неопределенности) , близки к размеру системы и поддерживают, соответственно, квантовые корреляции. Взаимодействие с окружением слабо и его величина легко контролируется.Таким образом, возможна провека квантовых когерентных оссциляций или туннельного эффекта на относительно больших, почти макроскпических системах. Все эксперименты проведенные до сих пор подтверждают выполнение Идеальной динамики, а не редукции в промежуточных состояниях.

2) Системы вблизи фазовых переходов II рода. Для таких систем велика длинна корреляции, сравнимая с размерами системы.

3) Может быть, какие-то типы живых процессов или их примитивные возможные прототипы.

**Часть 4. Определение Реальной Динамики,Идеальной Динамики, Непредсказуемой Динамики и Макроскопического Состояния. «Новая Динамика» Пригожина.**

Идеальной Динамикой мы условились называть исходные уравнения квантовой и классической механик. В реальности, кроме небольшого количества случаев, описанных выше, из-за невозможности полного самоописания системы или неустранимого
взаимодействия с внешним наблюдателем (средой) Идеальная Динамика экспериментально непроверяема и поведение системы становится, строго говоря, становится просто не предсказуемо.Появляется Непредсказуемая Динамика вместо Идеальной Динамики.

На практике, однако, большинство систем хорошо описывается и предсказывается законами физики. Как же это удается?!

Существует два основных фактора, приводящих к непредсказуемости:

1) Невозможность полного самоописания, если наблюдатель и среда включены в систему описания.Это накладывает ограничение на точное знание начальных условий движения.

2) Неконтролируемое взаимодействие внешнего наблюдателя или внешней среды с описываемой системой. Это накладывает ограничение на точное знание уравнений движения из-за неконтролируемого внешнего шума.

Есть, однако, решение этих трудностей. Оно состоит в замене полного описания системы на сокращенное через введение макроскопических переменных, являющимися некими функциями микропеременных.При этом здесь это понятие толкуется очень широко, например, знание скоростей и положений всех молекул с любой, но конечной точностью является также макроописанием системы.

**Потрясающий факт**, но для подавляющего числа реальных систем почти всегда существует (и не один!) набор макропеременных, при котором уравнения их движения становятся в очень широком диапазоне внешних шумов или погрешностей начальных условий НЕЗАВИСИМЫМИ (или почти НЕЗАВИСИМЫМИ) от величины и конкретного вида этих шумов или погрешностей начальных условий в течение промежутка времени меньшего половины времени возврата для периодических или почти периодических систем и даже бесконечному для хаотических систем или систем с бесконечным числом частиц или бесконечным размером.

Т.е. эта Реальная Динамика не зависит от погрешностей или внешних шумов, а зависит лишь от свойств самой системы, как и исходная Идеальная Динамика. Существует как минимум две причины делающие Реальную Динамику устойчивой к шуму: статистический закон больших чисел и дискретность квантовых переходов, обеспечивающую устойчивость химических связей [11].

Тут возникает очень важный вопрос, как выбрать макропеременные. Именно требование этой независимости от шума и накладывает ограничение на возможный выбор макропеременных. Например, кроме пары состояний «Шредингеровского кота» живой и мертвый мы можем выбрать в квантовой механике их полусумму и разность. Почему в реальности выбор остается лишь за парой кот живой или мертвый? Именно потому, что эта пара устойчива к малому шуму со стороны макроскопического окружения, в то время как их полусумма или разность распадаются на них даже при очень малом внешнем шуме (теорема Данери-Лойнжера-Проспери[30], [31] Daneri A., Loinger A., Prosperi G. M.)[14].

Возможны и другие ограничения на макропеременные, связанные, например, с желанием уменьшить их число или сделать их поведение более детерминированным.

Другим важным свойством Реальной Динамики является неоднозначность выбора самой этой динамики и набора макропеременных, который она описывает. По сути, мы разрабатываем некие новые фундаментальные законы для данного нового уровня описания, лишь опираясь на точную Идеальную Динамику, которая становится уже точно экспериментально не проверяемой, но конечный и точный выбор этих законов во многом определяется лишь их удобством для нас.

Все это позволяет заменить Непредсказуемую Динамику на предсказуемую Реальную Динамику макропеременных, которая получается через введение «огрубления» или «редукции» или массы аналогичных методов. Следует особо подчеркнуть, что в большинстве случаев Реальная Динамика не просто некое приближение Идеальной Динамики (как часто толкуется), разница между ними хоть реально и существует, но она просто НЕ наблюдаема экспериментально! Т.е. вопрос какая из этих динамик более верна становится просто бессмысленным.

Исходя из этой независимости результата Реальной Динамики, определяемой частотой измерения (редукции) при распаде частицы или переходе с одного энергетического уровня на другой, и выбирается интервал время между измерениями (редукции) таковым, чтобы в широкой окрестности этого интервала времени результаты не зависели от его точной величины, т.е. не слишком большим и не слишком малым, предотвращая тем самым парадокс котелка, который никогда не закипит.

В соответствии с двумя факторами непредсказуемости существует два типа Реальных Динамик:

1) Если наблюдатель и среда включены в систему описания.Это накладывает ограничение на точное знание начальных условий движения и приводит к Реальной Динамике не зависящей от них в широком интервале их значений. Это наиболее популярный тип Реальной Динамики, поскольку такая Динамика носит «объективный» характер и не зависит от внешних факторов, хотя, на самом деле, оба типа Реальных Динамик определяются лишь параметрами самой системы. К этому типу и относится и широко известная «Новая Динамика», разработанная Пригожиным [9], [12].

2) Если наблюдатель или среда НЕ включены в систему описания. Неконтролируемое взаимодействие внешнего наблюдателя или внешней среды с описываемой системой накладывает ограничение на точное знание уравнений движения из-за неконтролируемого внешнего шума и приводит к Реальной Динамике не зависящей от этого шума в широком интервале его значения и вида. Соответствует широко используемой и применяемой в квантовой механике «декогеренизации» квантовых систем, взаимодействующими с внешними «большими» макросистемами.

Остановимся подробнее на «Новой Динамике», разработанной Пригожиным [9], [12]. Она отличается от других аналогичных методов удачным выбором процедуры «огрубления». Дело в том, что большинство реальных замкнутых систем в классической механике - это системы с перемешиванием, где подавляющая часть траекторий экспоненциально неустойчивы. Их аналогом в квантовом случае являются системы бесконечного размера или системы с бесконечным числом частиц.Эти системы и рассматривает теория Пригожина. За бортом остается хоть и ограниченный, но важный класс классических периодических и почти периодических систем и почти все квантовые замкнутые системы, которые почти всегда обладают тем же свойством. Казалось бы, приведенные выше квантовые и классические системы рассматриваемые  Пригожиным принципиально различны в этом плане.В квантовых бесконечных системах время возврата бесконечно, а в классических хаотических системах имеет хоть и случайную, но конечную величину. Однако из-за неизбежных погрешностей самоизмерения в классических системах начальные условия размазаны в малой окрестности и благодаря случайным величинам времен возврата полный возврат системы, рассматриваемый как не одна точка в фазовом пространстве, а вся ее малая окрестность возможен тоже лишь за бесконечное время. Для индивидуальных же(одна точка) систем эти случайные времена возврата не могут быть самонаблюдаемы, внешний же реальный наблюдатель всегда внесет свою корректирующую погрешность.

Функция фазовой плотности обладает свойством сохранения фазового объема первоначальной малой окрестности. Поскольку для систем с перемешиванием близкие траектории в одном направлении экспоненциально расходятся из сохранения фазового объема следует, что в другом направлении они должны столь же быстро сходиться. В этом направлении и предлагается делать огрубление. Его максимальная величина определяется именно условием независимости или очень слабой зависимости макропеременных от величины огрубления, как и следует делать в реальной динамике.Эта процедура огрубления обладает замечательным свойством, выделяющим ее среди других: уравнения движения для огрубленной или не огрубленной функции фазовой плотности остаются эквивалентными при этой процедуре, в том смысле, что она перестановочна с процедурой огрубления т.е. безразлично, что сделать вначале: провести процедуру огрубления и использовать Пригожинские уравнения для получения конечной огрубленной функции фазовой плотности или использовать Идеальную Динамику, а в конце сделать это же огрубление.

При обращении скоростей, площадь охватываемая неогрубленной функцией фазовой плотности не меняется. Для обычных методов огрубления это свойство сохраняется, нарушается лишь обратимость уравнений движения, правда, как уже много раз говорилось, не регистрируемая экспериментально в реальных ситуациях. В уравнениях же Пригожина необратимость появляется из-за несимметричности процедуры огрубления при обратимости скоростей. Действительно, поскольку процедура огрубления идет в направлении сжатия, при обращении скоростей направлением сжатия становится направление расширения,
т.е. направлении протяженности «ветвей» функцией фазовой плотности, образующихся при «перемешивании», почти не меняя ее площади, в противоположность с необращенным вариантом, где направление огрубления ортогонально и перпендикулярно этим «ветвям». Энтропия огрубленной функции убывает с увеличением площади, охватываемым функцией фазовой плотности, следовательно, исходя из приведенных выше рассуждений, энтропия огрубленной функции, полученной из функции с обращением скоростей , выше энтропии исходной огрубленной функции, что приводит к росту энтропии и следовательно необратимости уравнений для огрубленной функции фазовой плотности. Какова величина огрубления ? Такова чтобы обеспечить реальную экспериментальную неразличимость «Новой» Реальной Динамики и Идеальной Динамики на протяжении бесконечного времени для описанных систем с перемешиванием. Это обстоятельство не было отмечено школой Пригожина.

Кстати, аналогичное огрубление может быть сделано и для периодических или почти периодических систем, правда, неразличимость Идеальная и «Новой» Динамики будет иметь место в этом случае лишь в течении времени равном половине периода возврата, что на самом деле и достаточно, как было показано выше, поскольку даже уже эти возвраты в реальных системах не наблюдаемы экспериментально. Это обстоятельство и не было замечено школой Пригожина. Какова величина огрубления ? Такова чтобы обеспечить реальную экспериментальную неразличимость «Новой» Реальной Динамики и Идеальной Динамики на протяжении полупериода времени для описанных периодических или почти периодических систем с перемешиванием в течении только полупериода .

Много споров возникает в связи с вопросом, что истинно Идеальная Динамика или Реальная «Новая Динамика» Пригожина[28]? Этот спор очень похож на спор о том, что вокруг чего вращается Земля вокруг Солнца или наоборот? На самом деле, по самому определению движения выбор остается за нами и определяется лишь красотой и нашим удобством, аналогично тому, чем в математической науке определяется выбор определений и теорем, и теория этого выбора, кстати, является еще неоткрытым континентом ее основ, в отличие от теоремы Геделя!
**(Примечание о т. Ферма: Кстати, результатом теоремы Геделя многие и объясняли труднодоказуемость т. Ферма, и итоговый результат оказался очень близок к этому утверждению. Т. Ферма получена как следствие гораздо более общей теоремы, чем она сама [13], и также включающей на порядок больше аксиом, чем теория натуральных чисел, на основе которых т. Ферма может быть сформулирована. Кстати сказать, было бы крайне любопытно получить от математиков не только 150 страничное доказательство, но и полный и вполне обозримый список аксиом математики, с указанием, какие из них использовались, а какие нет для этого доказательства. Аналогично этому, доказательство непротиворичивости арифметики получается**

**введением трансфинитной индукции и соответственным расширением списка аксиом. Правда, тут отсутствует обобщенная теорема, новая аксиома трансфинитной индукции вводится «руками». Что считать аксиомой, а что теоремой, опять таки, дело красоты и удобства. А красота спасет мир!)**

Аналогично этому, разница между «Новой» и Идеальной Динамикой в большинстве реальных ситуаций экспериментально ненаблюдаема и выбор по большому счету произволен. В тех редких случаях «больших» систем, когда Идеальная Динамика точно проверяема она всегда пока что побеждает. Может быть исследуемые системы пока не достаточно большие? Ответ за дальнейшими экспериментами. Что окажется верным и какая Динамика из очень многих возможных одержит победу можно лишь гадать. Ситуация тут похожа на Великие Струнные Теории и Великие Объединения: покачто лишь догадки, а до эксперимента, который должен дать ответ может быть сотни лет, если не поможет Космос или Пришельцы, впрочем и в теории гравитации Энштейна, которая почему-то считается несомненной истиной (вот она,сила авторитета!), мы имеем пока в эксперименте лишь намеки на истину **(вспомним теории гравитации Мильгрома и Логунова и загадочное темное вещество)** . Сколько же ждать тут? Только Бог знает.

Закончим эту часть важным замечанием. Наряду с системами, описываемыми Реальной Динамикой или Идеальной Динамикой могут существовать системы, которые при попытке описать их подробно и детально отвечают Непредсказуемой Динамике, т.е. их детальное описание хоть и теоретически возможно, но не проверяемо экспериментально и не реализуемо на практике. Возможно к такому типу систем и относятся живые системы.

**Часть 5. Жизнь и смерть.**

Отметим с самого начала, что если предыдущие части носили более или менее строгий характер, данная в силу очевидных причин носит более гипотетический характер и является скорее набором гипотез.

Будем исходить здесь из положения, что жизнь полностью соответствует законам физики.

Что такое жизнь и смерть с точки зрения физики?

Есть ли у живой материи некие свойства не совместимые с физикой?

Чем живые системы отличаются от неживых с точки зрения физики?

Когда у живых систем появляется сознание и свобода воли
с точки зрения физики?

Жизнь определяется, обычно, как особая высокоорганизованная форма существования органических молекул, обладающая способностью к обмену веществ, размножению, адаптации, движению, реакцией на внешние раздражители, способностью к самосохранению в течении долгого времени или даже повышению уровня самоорганизации. Это верное, но слишком узкое определение: многие из живых систем обладают лишь частью из этих свойств, некоторые из них присущи и неживой материи, вполне возможны и неорганические формы жизни.

Первую попытку описать жизнь с точки зрения физики дал Шредингер [11]. В своей работе он определил жизнь как апериодический кристалл, т.е. высокоупорядоченную (и, соответственно, обладающую низкой энтропией и

«питающуюся» негоэнтропией из окружения, т.е. принципиально открытую систему), но не основанную на простом повторении, в отличие от кристалла, форму материи и привел две причины делающие Реальную Динамику живых систем устойчивой к их внутреннему и внешнему шуму: статистический закон больших чисел и дискретность квантовых переходов, обеспечивающую устойчивость химических связей. Сам принцип действия живых организмов он уподобляет часам: и там и там возникает «порядок из порядка» несмотря на высокую температуру.

В своей работе советский биофизик Бауэр [12] определил, что не только высокая упорядоченность (и, соответственно, низкая энтропия) проявляются не только в неравновесности распределения веществ в живой материи, но и сама структура живой материи является низкоэнтропийной и сильно неустойчивой. Эта неустойчивая структура не только поддерживается за счет процесса обмена веществ, но и является их катализатором. Это предположение верно лишь частично, например белки или вирусы сохраняют свою структуру и в кристаллической форме, но их низкоэнтропийные и сильно неустойчивые модификации и сочетания внутри живой материи обладают этим свойством. С течением времени, тем не менее, происходит постепенная деградация структуры, что и приводит к неизбежности смерти и необходимости размножения для сохранения жизни. Т.е. процесс обмена веществ лишь очень сильно замедляет распад сложной структуры живой материи, а не поддерживает ее все время неизменной. Экспериментальные результаты, приведенные Бауэром, подтверждают выделение энергии и соответственно увеличение энтропии в процессе автолиза, т.е. распада живой материи. Он видит его причину на первой стадии процесса в неустойчивости самой исходной структуры без поддерживающего ее обмена веществ и на второй стадией процесса в действии протолитических (разлагающих) ферментов, освобождающихся или появляющихся при автолизе. Наличие этой избыточной структурной энергии Бауэр и считал неотемлимой характеристикой жизни.

Во всех этих работах дано рассмотрение отдельного живого организма, в то время как жизнь, как совокупность всех организмов в целом (биосфера) может быть рассмотрено и определено. Сюда же относится и вопрос о происхождении и источнике жизни. Наиболее полный и современный ответ на эти вопросы с точки зрения физики был дан в работе Элицура [18]. В ней он рассматривает источник жизни как ансамбль саморазмножающихся молекул. Проходя через сито Дарвиновского естественного отбора, жизнь накапливает в своих генах информацию (или скорее знания в определении, данном Элицуром) об окружающей среде, повышая, тем самым, уровень своей организации (негоэнтропии) в соответствии со вторым началом термодинамики. Ламаркизм в его слишком прямолинейной формулировке приведен в противоречие с этим законом физики.Взгляд на широкий спектр работ в этой области отражен в статье. К недостаткам работы относятся:

1) Рассмотрение верно для жизни в целом, как явления, но не для отдельно взятого живого организма.

2) Предложенное доказательство отвергает лишь слишком грубую, прямолинейную модель Ламаркизма, в то время как есть много гипотез и опытов, иллюстрирующих возможность реализации его элементов даже в реальной жизни[32].
3) За самоорганизующимися диссипативными системами, предложенными Пригожиным, например ячейками Бернара, отрицается свойство адаптации, в отличие от живых организмов. Естественно, их адаптивные способности не сравнимы с живыми системами, но в зачаточной форме, тем не менее, существуют. Так, например, ячейки Бернара,меняют свою геометрию или даже исчезают, как функция разницы температур между нижним и верхним слоем жидкости.Это и есть примитивная форма адаптации.

Равновесный ансамбль, находящийся в равновесии с термостатом, в квантовой механике в энергетическом представлении описывается диагональной матрицей плотности.Если есть параметры, кроме энергии, необходимые для полного описания системы, величинами диагональных членов им соответствующие равны. Аналогично этому, в классической механике, в равновесии отсутствуют корреляции между молекулами, являющимися аналогами недиагональных элементов матрицы плотности.

Таким образом нарушение равновесия проявляется двояко: в неравновесном распределении диагональных элементов и в неравенстве нулю недиагональных элементов (что соответствует ненулевым корреляциям в классической механики), причем эти корреляции много более неустойчивы и гораздо быстрее затухают, чем отклонение диагональных элементов от равновесных величин.

Поскольку жизнь определена Бауэром как самоподдерживающаяся за счет движения и обмена веществ сильная неустойчивость, мы можем предположить, что большая часть этой неустойчивости проистекает из этих сильно неустойчивых корреляций (в квантовой механике недиагональных элементов), которые живые системы стремятся поддержать и сохранить в течении времени много большем их времени релаксации. В неживых системах это достигается просто изоляцией системы, в живых же открытых системах, активно взаимодействующих с окружением это достигается их внешним и внутренним движением и метаболизмом. Этим живые системы подобны изолированным системам, в которых корреляции пассивно сохраняются за счет этой изоляции, в живых же открытых системах они поддерживаются за счет их активного взаимодействия с окружением. Следует отметить, что живые системы поддерживают корреляции как между внутренними элементами, так и корреляции с окружением.

Введем понятие псевдоживых физических систем. Будем называть таковыми простые физические системы иллюстрирующие в зачаточной форме некие действительные или предполагаемые свойства живых систем. Так, рост кристаллов моделирует способность живых систем к размножению. Кстати, анализ этих систем позволяет найти слабое место в аргументах Вигнера [6] , [27], видящем противоречие между способностью к размножению и квантовой механикой.

Другой пример - это квантовые изолированные системы, демонстрирующие свойство сохранения корреляций, аналогичных поддержанию сильной неустойчивости в живых системах,связанной с сохранением корреляций или недиагональных элементов матрицы плотности.

Правда, это сохранение пассивно. Активное замедление времен релаксации недиагональных элементов матрицы плотности, более близкое к методам поддержания корреляций в живых системах, достигается в таких открытых системах, таких как микромазеры [25], которые являются еще одним примером псевдоживых систем. Диссипативные системы иллюстрируют свойства открытых живых систем к поддержанию низкой энтропии и примитивной адаптации к изменению условий окружающей среды.

Кстати, определение жизни, как систем, способствующих сохранению корреляций в противовес внешнему шуму, хорошо объясняет загадочное молчание КОСМОСА, т.е. отсутствие сигналов от других разумных миров. Вселенная произошла из единого центра (Большой Взрыв) и все ее части коррелированны, жизнь лишь поддерживает эти корреляции и существует на их основе. Поэтому процессы возникновения жизни в различных частях скоррелированны и находятся на одном уровне развития, т.е. сверхцивилизаций, способных достичь Земли пока просто нет. Эффектами дальних корреляций можно объяснить и часть поистине чудесных проявлений человеческой интуиции и парапсихологических эффектов. Причем здесь необязательна квантовая механика, подобные корреляции присущи и классической механике, имеющей аналоги недиагональных элементов матрицы плотности. Подобные корреляции часто ошибочно увязывают лишь с квантовой механикой.

Следующий вклад в понимание жизни сделал Бор [26]. Он обратил внимание, что полное измерение состояния системы вносит в квантовой механике неизбежные искажения в поведение системы, чем возможно и объясняется принципиальная непознаваемость жизни. Критика этих взглядов Бора Шредингером [19] не состоятельна. Она основана на том, что полное измерение состояния системы возможно и в квантовой механике, просто оно отлично от классического- оно вероятностно. Проблема не в том, что такое измерение невозможно. Истинная проблема заключается в том, что подобное измерение меняет дальнейшее поведение системы, в отсутствии измерения оно было бы иным [20]. Измерение нарушает тонкие корреляции между частями системы, меняя ее поведение. Это относится не только к  квантовой механике, но и к классической механике, где между реальными системами существует конечное взаимодействие.

Псевдофизическими системами иллюстрирущими свойство измерения нарушать динамику систем являются оссцилирующие квантовые почти изолированные системы, изменяющиеся по схеме: A -> сумма A и B -> B -> разница A и B-> A, где A и B - состояния системы. Измерение в каком состоянии находится система A или B нарушает состояния их суммы или разницы, меняя реальную динамику системы и уничтожая корреляции ( не диагональные элементы матрицы плотности) между A и B в этих состояниях [10].

Успехи молекулярной генетики не опровергают этой точки зрения. Построение Реальной Динамики жизни в принципе возможно. Действительно, живые системы - это открытые системы, активно взаимодействующие со случайным окружением. Внешний наблюдатель взаимодействует с ними обычно много слабее и не может вызвать принципиальное изменение в их поведении. Однако попытка понять и предсказать жизнь слишком подробно и детально нарушит сложные и тонкие корреляции, сохраняемые жизнью, и приведет к Непредсказуемой Динамике живых систем и давая эффект, предсказанный Бором. Возможно, особо тонкая человеческая интуиция и некоторые парапсихологические эффекты и лежат в этой области Непредсказуемости. То что они могут лежать только в этой узкой области на грани постижимости точной наукой и не дает естественному отбору усилить эти свойства, так и не дает нам возможности постичь полностью эти явления средствами науки [21], [22].

**Часть 6. Заключение.**

Приведенная статья не является лишь философским абстрактным построением. Не понимание этих основ приводит к ошибкам. Примерами могут служить ошибки сделанные в теории полюсов в задачах движения фронта пламени и рост «пальца» на поверхности раздела жидкостей.

Севашинский[23] утверждал, что Идеальная Динамика полюсов приводит к ускорению фронта пламени, и это ускорение не вызвано шумом, поскольку оно не меняется при уменьшении шума и зависит лишь от свойств самой системы. Но ведь также Реальная Динамика, связанная с шумом, не зависит от него в широком интервале значений.

Танвир [24] нашел различие в росте «пальца» в теории и численных экспериментах, не поняв, что эта разница связана с численным шумом, приводящим к новой Реальной Динамике. Это лишь два рядовых примера, взятых из повседневной практики автора статьи, а встретить их можно много. Изложенные в этой статье результаты необходимы для понимания основ нелинейной динамики, термодинамики и квантовой механики.

**Литература.**

**Приложение А. Матрица плотности**

Рассмотрим пучок из $N_a$ частиц , приготовленных в состоянии $|\chi_a\rangle$ , и еще один независимый от первого пучок из $N_b$ частиц, приготовленных в состоянии $|\chi_b\rangle$ . Для описания объединенного пучка введем оператор ρ смешанного состояния, определяемый выражением

$$\rho=W_a|\chi_a\rangle\langle\chi_b| + W_b|\chi_a\rangle\langle\chi_b|$$

где $W_a=N_a/N$, $W_b=N_b/N$, $N=N_a+N_b$

Оператор ρ называют оператором плотности или статистическим оператором. Он описывает способ приготовления пучков и тем самым содержит всю информацию о полном пучке. В этом смысле смесь полностью определена своим оператором плотности. В частном случае чистого состояния $|\chi\rangle$ оператор плотности дается выражением

$$\rho=|\chi\rangle\langle\chi|.$$

Обычно более удобно записывать оператор ρ в матричной форме. Для этого выберем набор базисных состояний ( обычно $|+1/2\rangle$ и $|-1/2\rangle$) и разложим состояния $|\chi_a\rangle$ и $|\chi_b\rangle$ по этому набору согласно соотношению

$$|\chi_a\rangle=a_1^{(a)}|+1/2\rangle+a_2^{(a)}|-1/2\rangle,$$
$$|\chi_b\rangle=a_1^{(b)}|+1/2\rangle+a_2^{(b)}|-1/2\rangle.$$

В представлении состояний $|\pm1/2\rangle$ :

$$|\chi_a\rangle=\begin{pmatrix} a_1^{(a)} \\ a_2^{(a)} \end{pmatrix}$$

$$|\chi_b\rangle=\begin{pmatrix} a_1^{(b)} \\ a_2^{(b)} \end{pmatrix}$$

а для сопряженных состояний -

$$\langle\chi_a|=(a_1^{(a)*}, a_2^{(a)*}),$$
$$\langle\chi_b|=(a_1^{(b)*}, a_2^{(b)*}).$$

Применяя правила умножения матриц, получим для «внешнего произведения»

$$|\chi_a\rangle\langle\chi_a| = \begin{pmatrix} a_1^{(a)} \\ a_2^{(a)} \end{pmatrix} (a_1^{(a)*},\ a_2^{(a)*}) = \begin{pmatrix} |\,a_1^{(a)}\,|^2 & a_{{}_1 1}^{(a)} a_2^{(a)*} \\ a_1^{(a)*} a_2^{(a)} & |\,a_2^{(a)}\,|^2 \end{pmatrix}$$

и аналогичное выражение для произведения $|\chi_b\rangle\langle\chi_b|$ . Подставляя эти выражения в оператор плотности, находим
матрицу плотности

$$\rho = \begin{pmatrix} W_a\left|a_1^{(a)}\right|^2 + W_b\left|a_1^{(b)}\right|^2 & W_a a_1^{(a)} a_2^{(a)*} + W_b a_1^{(b)} a_2^{(b)*} \\ W_a a_1^{(a)*} a_2^{(a)} + W_b a_1^{(b)*} a_2^{(b)} & W_a\left|a_2^{(a)}\right|^2 + W_b\left|a_2^{(b)}\right|^2 \end{pmatrix}$$

Поскольку при выводе этого выражения были использованны базисные состояния $|\pm 1/2\rangle$, полученное выражение называют матрицей плотности в $\{\ |\pm 1/2\rangle\}$-представлении.

В заключение несколько слов о статистической матрице $P_0$, обладающей замечательными свойствами. Мы знаем, что в классической статистической термодинамке все возможные макроскопические состояния системы рассматриваются как априори равновероятными (другим словами, они считаются равновероятными, если нет каких-либо сведений о значении полной энергии или о контакте с термостатом, поддерживающим постоянную температуру системы, и т.д.). По аналогии с этим в волновой механике все состояния системы, определяемые различными функциями, образующие полную систему ортонормированных функций, можно априори предполагать равновероятными. Пусть $\varphi_1,\ldots,\varphi_k$,- такая система базисных функций $\varphi_k$; зная, что система характеризуется смесью состояний $\varphi_k$, в отсутствие какой-либо другой информации можно считать, что статистическая матрица системы имеет вид

$$P_0 = \sum_k p P_{\varphi_k}\ , \text{ где } \sum_k p = 1\ ,$$

т.е., что $P_0$ - статистическая матрица такого смешанного состояния, для которого все веса равны между собой. Принимаем $\varphi_k$ за базисные функции, матрицу $P_0$ можно представить в виде

$$(P_0)_{kl} = p\delta_{kl}$$

Если статистическое состояние ансамбля систем в начальный момент времени характеризуется матрицей $P_0$ и если во всех системах ансамбля провести измерение одной и той же величины A, то статистическое состояние ансамбля будет по прежнему характеризоваться матрицей $P_0$.
Уравнения движения для матрицы плотности $\rho$

$$i\frac{\partial\rho_N}{\partial t} = L\rho_N$$

где L - линейный оператор:
$L\rho = H\rho - \rho H = [H,\rho]$,
где H - оператор энергии системы.

Если A – оператор некой наблюдаемой величины,

Среднее значение этой величины может быть найдено следующим образом:
$<A>=\mathrm{tr}A\rho$

**Приложение B** Редукция матрицы плотности и теория измерения.
Пусть при измерении некоторого объекта мы «четко различаем» состояния $\sigma^{(1)}$, $\sigma^{(2)}$, ... . Производя измерения над объектом, находящимся в этих состояниях, мы получаем числа $\lambda_1, \lambda_2$, ... . Начальное состояние измерительного прибора обозначим через a. Если измеряемая система сначала находилась в состоянии $\sigma^{(v)}$, то состояние полной системы «объект измерения плюс измерительный прибор» до того, как они вступили во взаимодействие будет определяться прямым произведением $a \times \sigma^{(v)}$ . После измерения:

$$a \times \sigma^{(v)} \rightarrow a^{(v)} \times \sigma^{(v)}$$

Пусть теперь начальное состояние будет не «четко различимым», а произвольной комбинацией $\alpha_1\,\sigma^{(1)} + \alpha_2\,\sigma^{(2)}+...$
таких состояний. В этом случае из линейности квантовых уравнений следует

$$a \times [\Sigma\alpha_v\sigma^{(v)}] \rightarrow \Sigma\alpha_v[a^{(v)} \times \sigma^{(v)}]$$

В итоговом состоянии, возникающем в результате измерения,
существует статистическая корреляция между состоянием объекта и состоянием прибора: одновременное измерение у системы «объект измерения плюс измерительный прибор» двух величин ( первой- подлежащий измерению характеристики исследуемого объекта и второй- положение стрелки прибора) всегда приводит к согласующимся результатам. Вследствие этого одно из названных измерений становится излишним - к заключению о состоянии объекта измерения можно прийти на основании наблюдения за прибором.
Результатом измерения является не вектор состояния, представляемый в виде суммы, стоящей в правой части найденного соотношения, а так называемая смесь, т.е. один из векторов состояния вида

$$a^{(v)} \times \sigma^{(v)}$$

и что при взаимодействии между объектом измерения и измерительным прибором это состояние возникает с вероятностью $|\alpha_v|^2$. Данный переход называется редукцией волнового пакета и соответствует переходу матрицы плотности из недиагонального $\alpha_v\alpha_\mu^*$ в диагональный вид $|\alpha_v|^2\,\delta_{v\mu}$. Этот переход не описывается квантовыми уравнениями движения.

**Приложение C** Функция фазовой плотности
В данный момент времени t состояние системы N одинаковых частиц с точки зрения классической механики определяется заданием значений координат $\mathbf{r}_1, \ldots, \mathbf{r}_N$ и импульсов $\mathbf{p}_1, \ldots, \mathbf{p}_N$ всех N частиц системы. Для краткости будем использовать обозначение

$$x_i=(\mathbf{r}_i,\mathbf{p}_i) \quad (i=1, 2, \ldots, N)$$

для совокупности значений координат и компонент импульса отдельной частицы и обозначение

$$X= (x_1, \ldots, x_N)=(\mathbf{r}_1, \ldots, \mathbf{r}_N\,, \mathbf{p}_1, \ldots, \mathbf{p}_N)$$

для совокупности значений координат и значений импульсов всех частиц системы. Такое пространство 6N переменных называют 6N-мерным фазовым пространством. Чтобы определить понятие функции распределения состояний системы, рассмотрим набор одинаковых макроскопических систем - ансамбль Гиббса. Для всех этих систем условия опыта одинаковы. Поскольку, однако, эти условия определяют состояние системы неоднозначно, то разным состояниям ансамбля в данный момент времени t будут соответствовать разные значения X.
Выделим в фазовом пространстве объем dX около точки X. Пусть в данный момент времени t в этом объеме заключены точки, характеризующие состояния dM систем ансамбля из их полного числа M. Тогда предел отношения этих величин

$$\lim_{m\to\infty} dM/M = f_N(X,t)dX$$

и определяет фазовую функцию плотности распределения
в момент времени t.

$$\int f_N(X,t)dX=1$$

Уравнение Лиувилля для функции фазовой плотности
Можно записать в виде

$$i\frac{\partial f_N}{\partial t} = Lf_N = \{H, f_N\}$$

где L - линейный оператор:

$$L = -i\frac{\partial H}{\partial p}\frac{\partial}{\partial x} + i\frac{\partial H}{\partial x}\frac{\partial}{\partial p},$$

где H - энергия системы.

**Приложение D Огрубление функция фазовой плотности и гипотеза молекулярного хаоса**.

Огрублением функции плотности будем называть ее замену на приближенную величину.
Например

$$f_N^*(X,t)=\int_{(y)} g(X-Y)f_N(Y,t)$$

где

$$g(X)=1/\Delta\, D(X/\Delta)$$

$D(x)= 1$ for $|X|<1$
$D(x)= 0$ for $|X|\geq 1$

Другой пример огрубления - «гипотеза молекулярного хаоса»

Замена двухчастичной функции распределения на произведение одночастичных функций
$f(x_1,x_2,t) \to f(x_1,t)f(x_2,t)$

**Приложение E Определения энтропии.**
В качестве определения энтропии укажем выражение

$S=-k \int_{(X)} f_N(X,t) \ln f_N(X,t)$

Для квантовой механики через матрицу плотности

$S=-k\, tr\, \rho \ln \rho$ [29]

где tr - след матрицы.

Эти энтропии не меняются при обратимой эволюции. Чтобы получить меняющиеся энтропии
в качестве $f_N$ или $\rho$ используются их огрубленные величины.

**Приложение F Новая динамика Пригожина.**
Позвольте ввести линейный оператор $\Lambda$ на функции $\rho$ фазовой плотности или матрицы плотности

$\tilde{\rho} = \Lambda^{-1}\rho$
$\Lambda^{-1}\, 1=1$
$\int \tilde{\rho} = \int \rho$
$\Lambda^{-1}$ сохраняет положительность.
Такой, что функция $\tilde{\rho}$ обладает свойством:для функции $\Omega$ определенной как
$\Omega = tr\tilde{\rho}^{+}\tilde{\rho}$ или $\Omega = tr\tilde{\rho}\, \ln\tilde{\rho}$
имеем $d\Omega/dt \leq 0$
Уравнение движения для $\rho$

$$\frac{\partial\tilde{\rho}}{\partial t} = \Phi\tilde{\rho}$$

где $\Phi = \Lambda^{-1} L\Lambda$
$\Phi$-необратимая марковская полугруппа
$\Lambda^{-1}(L)=\Lambda^{+}(-L)$
Для функции фазовой плотности оператор $\Lambda^{-1}$ соответствует
огрублению в направлении сжатия фазового объема. Для квантовой механики такой оператор может быть найден лишь для бесконечного объема или бесконечного числа частиц. Введем в квантовой механике проекционный оператор P, обнуляющий недиагональные элементы матрицы плотности. Оператор $\Phi$ и базисные вектора матрицы плотности выбираются таким образом, чтобы сделать оператор $\Phi$ перестановочным с оператором P:

$$\frac{\partial \mathrm{P}\tilde{\rho}}{\partial t} = \mathrm{P}\Phi\tilde{\rho} = \Phi\mathrm{P}\tilde{\rho}$$

**Приложение G** Невозможность **самонаблюдения.**
Предположим, что существует мощная вычислительная машина способная предсказать будущее свое и окружения на основе расчета движения всех молекул.Пусть ее предсказания - выкатывание черного или белого шара из некого устройства. Устройство выкатывает белый шар, если машина предсказывает черный и наоборот черный шар при предсказании белого. Ясно, что предсказания машины всегда не верны, что доказывает невозможность самонаблюдения и саморасчета.
**Приложение H** Соответствие **квантовой и классической механик. Таблица.**

| Квантовая механика | Классическая механика |
|---|---|
| Матрица плотности | Функция фазовой плотности |
| Уравнение движения для матрицы плотности | Уравнение Лиувилля |
| Редукция волнового пакета | Огрубление функции фазовой плотности или гипотеза молекулярного хаоса |
| Неустранимое воздействие наблюдателя или окружения, описываемое редукцией | Теоретически бесконечно малое, но в реальности конечное взаимодествие системы с наблюдателем или окружением |
| Корреляции между скоростями и положениями молекул в разных частях системы | Ненулевые недиагональные элементы матрицы плотности |